\documentclass[11pt]{article}

\title{Practical LR Parser Generation}
\author{Joe Zimmerman \vspace*{-0.1em} \\ \small{\texttt{jzim@cs.stanford.edu}}}
\date{}

\usepackage{amsfonts}
\usepackage{amsmath}
\usepackage{amssymb}
\usepackage{amsthm}
\usepackage{array}
\usepackage{enumerate}
\usepackage{enumitem}
\usepackage[left=1in,right=1in,top=1in,bottom=1in,columnsep=0.25in]{geometry}
\usepackage[hidelinks]{hyperref}

\newtheorem{theorem}{Theorem}[section]
\newtheorem{lemma}[theorem]{Lemma}

\newtheorem{definition}[theorem]{Definition}

\newtheorem{construction}[theorem]{Construction}

\setlist{listparindent=\parindent,parsep=0pt}

\newcommand{\esm}[1]{\ensuremath{#1}}
\newcommand{\mr}[1]{\esm{\mathrm{#1}}}

\newcommand{\mc}[1]{\esm{\mathcal{#1}}}

\newcommand{\eps}{\esm{\varepsilon}}

\newcommand{\ra}{\esm{\rightarrow}}

\newcommand{\GG}{\esm{\mc{G}}}

\newcommand{\genstar}{\esm{\overset{*}{\Rightarrow}}}

\newcommand{\srt}{\mbox{\esm{\,\dashv\,}}}

\newcommand{\fst}{\esm{\mr{First}}}
\newcommand{\fstgen}{\esm{\mr{Gen}}}

\begin{document} 

\maketitle

\begin{abstract}
    The parsing problem, taking a string of characters and producing an abstract syntax tree,
    is a fundamental building block in modern compilers.
    For industrial programming languages, parsing is a surprisingly involved task.
    Approaches to generate parsers automatically have been known for many years,
    beginning with Knuth's seminal work on canonical LR parsing in 1965,
    and refined by DeRemer in 1969 to the restricted LALR class of grammars
    (which serve as the basis for automatic parser generation tools such as {\texttt{yacc}}
    and {\texttt{bison}}).
    However, the prevailing consensus is that automatic parser generation is not practical
    for real programming languages:~LALR parsers---and even the more general LR parsers,
    in many cases---are
    considered to be far too restrictive in the grammars
    they support, and moreover, LR parsers are often considered too inefficient in practice.
    As a result, virtually all modern languages use recursive-descent parsers
    written by hand, a lengthy and error-prone process that dramatically increases
    the barrier to new programming language development.

    In this work we demonstrate that, contrary to the prevailing consensus, we can
    have the best of both worlds:~for a very general, practical class of grammars---a strict superset
    of Knuth's canonical LR---we can generate parsers automatically, and such that the
    resulting parser code, as well as the generation procedure itself, is highly efficient.
    This advance relies on several new ideas, primarily:~(i)
    an improved optimization procedure for lookaheads in LR automata;
    (ii) a grammar transformation, reminiscent of continuation-passing style (CPS),
    which defers shift/reduce decisions to the end of processing for compound rules;
    (iii) a refinement of the LR paradigm to include recursive-descent actions;
    (iv) an extension of context-free grammars to include per-symbol attributes
    (e.g., ``expression in a type context vs.~a value context'');
    and (v) an extension of canonical LR parsing, which we refer to as XLR,
    which endows shift/reduce parsers with the power of bounded nondeterministic choice.

    With these ingredients, we can automatically generate efficient parsers
    for virtually all programming languages
    that are intuitively easy to parse---a claim we support experimentally,
    by implementing the new algorithms in a new software tool called \texttt{langcc}~\cite{langcc}
    and running them on syntax specifications for Golang~1.17.8 and Python~3.9.12.
    The tool handles both languages automatically,
    and the generated code, when run on standard codebases,
    is 1.2x faster than the corresponding hand-written parser for Golang,
    and 4.3x faster than the CPython parser, respectively.
\end{abstract}

\newpage

\section{Introduction}

\subsection{General parsing}

We begin with a brief exposition of the parsing problem in its full generality.
Parsing takes as input a context-free grammar
$\GG$ and some input string of length $n$, and produces a parse tree for the string
according to $\GG$, if one exists.
General parsing can be done in time $O(n^3)$, neglecting factors of~$|\GG|$,
by a standard bottom-up dynamic programming approach.
Valiant's algorithm~\cite{Val75} improves this result to $O(n^\omega)$, where $\omega$ denotes the matrix
multiplication exponent.
The algorithms of Earley~\cite{Earley}, Tomita~\cite{Tomita} and Leo~\cite{Leo}
(``ETL'') also build a dynamic programming
table, but do so left-to-right rather than bottom-up, and hence only generate
parses for substrings that are consistent with some left-to-right partial parse
of the entire input up to that point.
As a result, while the ETL algorithms still require $\Omega(n^3)$ time
for general grammars, they achieve much better performance on various specific
subclasses of grammars.

\subsection{Unambiguous grammars}

Virtually all programming language grammars of practical interest are {\em unambiguous},
meaning that for every reachable symbol $X$ of the grammar
and every string $x$, there is at most one parse tree $X \genstar x$.
Of the ETL algorithms, Earley's and Leo's both run in time $O(n^2)$ on unambiguous grammars,
despite requiring $O(n^3)$ time in the general case.
Naively, we might hope that the class of grammars for which one could automatically
generate efficient parsers would turn out to be precisely the class of unambiguous grammars.

Unfortunately, it is currently unknown whether unambiguous grammars can be parsed in
sub-quadratic time.
Abboud et al.~\cite{ABW15} give a reduction showing that general grammars cannot be parsed in
time $o(n^\omega)$,
unless there is a breakthrough in searching for cliques in graphs.
However, clearly this reduction does not extend to the case of unambiguous grammars,
since it is already known how to parse these in time $O(n^2)$ via ETL.
In fact, as far as we know, it could be possible to parse
every unambiguous grammar in $O(n)$ time.
Intuitively, we conjecture a pessimistic outcome;
however, we observe that we currently do not even have an example of a specific unambiguous
grammar for which no linear-time parsing algorithm is known.
This gap is one of the most significant open problems in the study of parsing,
as it is of not only theoretical but potentially great practical importance.
For the present, however, we require practical parsers to run in $O(n)$ time,
and hence we must accept some limitations beyond just unambiguity.

\subsection{Practical parsers}

For industrial programming languages, the gold standard is the hand-written recursive-descent parser,
which typically runs in time $O(n)$ with a very small constant factor, independent of $|\GG|$.\footnote{The
    lack of dependence on $|\GG|$,
    which might seem impossible a priori, is achieved by encoding the grammar into the structure
    of the code itself (for hand-written parsers), or into a series of branch tables (for automatically
    generated parsers).}
This means that, in order for automatic parser generation to be competitive with this approach,
it is not sufficient merely to require the input grammar to be unambiguous.
In fact, even Leo's algorithm,
which runs in time $O(n)$ on all grammars in the class LR($k$),
poses efficiency problems for competitive automatic parser generation,
due to the associated constant factors.
Thus, for algorithms which are competitive with hand-written parsers in practice,
we look to the highly efficient online (left-to-right) family of parsing algorithms,
which typically require grammars of a more restricted form.

Of these online left-to-right algorithms,
the primary categories are top-down (``LL($k$)'') and bottom-up (``LR($k$)'').
In top-down parsing, we require the online parser, with $k$ symbols of lookahead, to emit the name of each production
when its occurrence {\em begins} in the input;
in bottom-up parsing,
it is emitted when its occurence {\em ends}.
It is fairly clear that top-down parsing is not sufficiently general for practical
programming languages:~even
in simple cases such as binary operator-precedence grammars,
one does not know whether one is parsing the production ``$E \ra E + E$'' or the production ``$E \ra E \times E$''
until after seeing the entirety of the first occurrence of $E$.
While there are transformations of the input grammar that can make it more amenable
to top-down parsing, many reasonable languages admit no LL($k$) grammar for any $k$
(e.g., the well-known ``dangling-else'' grammar).
Moreover, even in cases where such transformations succeed,
they often introduce significant overhead and drastically change the semantics of the grammar.
Thus, in order to achieve automatic parser generation for a practical family of languages,
we will confine our attention to bottom-up (LR-style) parsing.

\subsection{Practical classes of grammars}

Within the category of online bottom-up parsing, there remains the question
of what restrictions we may place on the input grammar,
beyond simple unambiguity.
Knuth's LR($k$) is a reasonable starting point, which we proceed to refine as follows.
First, we address the question of which direction to refine the class of LR grammars:~i.e., 
whether, for efficiency, we aim for a smaller class such as SLR or LALR grammars~\cite{DeRemer};
or, for generality, we aim for a {\em larger} class.
It is well-known that SLR and LALR, while being highly efficient in their implementation,
fail to capture many natural constructs for which it is easy to write recursive descent parsers by hand.
Thus, in order to be competitive with hand-written recursive-descent parsers,
we should aim for a class of grammars {\em at least} as general as LR.
On the other hand,
conventional wisdom is that even LR($1$) is
often too inefficient to be practical.

In this work, we argue that both of these claims are incorrect:~LR($1$) parsing
can be practical, with certain optimizations; and, moreover, even LR($1$) parsing
is not sufficiently general to handle languages that are intuitively ``easy to parse''
via hand-written recursive descent.
To support these claims, we reason as follows.
    
\ \\
\noindent \textbf{LR parser generation can be practical.}\ \ 
Traditionally, canonical LR is seen as impractical
due to the proliferation of states in the generated automata.
Specifically, the naive procedure calls for
NFA states $(X \ra \alpha \cdot \beta,\,\lambda)$
for every dotted production $X \ra \alpha \cdot \beta$
and for every possible $k$-follow\footnote{Traditionally, the term ``lookahead''
is used to refer to both (a) the next $k$ tokens in the input string,
and (b) the $k$ tokens that follow a given production according to
a vertex in the LR automaton.
To avoid confusion,
in this work we use the term ``lookahead'' to refer only to the former,
and the term ``$k$-follow string'' (or ``$k$-follow set'')
to refer to the latter.}
string $\lambda$.
Since an industrial programming
language might easily have over $1000$ productions
and over $50$ symbols, this essentially rules out practicality of LR($k$) for $k > 1$,
and in some cases even $k = 1$ becomes slow enough to be questionable
for a modern compiler toolchain.

In this work, we propose an optimized LR NFA construction procedure that
drastically reduces the number of states required, which in turn improves both the
efficiency and the debuggability of the resulting automata.
At a high level, the construction proceeds by first building the LR($0$) automaton
(i.e., an automaton with no $k$-follow strings);
determining what $k$-follow strings are possible for each state,
and, of these, which are involved in simple LR conflicts;
and propagating these conflicting strings through the automaton to determine a partition
of each state's $k$-follow set.
(We defer the full details of the construction procedure to Section~\ref{sec:optaut}.)

Empirically, we find that even in languages that are relatively difficult to parse,
such as Golang, there are relatively few $k$-follow strings that are involved in simple LR conflicts.
Moreover, when these strings are propagated backwards through the automaton,
the partitions are often completely annihilated by other productions.
(For instance, if $W \ra X \cdot Y Z$ and $Y \ra \cdot \, A B$ are dotted productions,
then a conflicting $k$-follow string on $(Y \ra \cdot \, A B,\,\lambda)$
will not induce a partition of the $k$-follow strings on $W \ra X \cdot Y Z$ unless $Z$ generates a string
of length less than $k$---i.e., if $k = 1$, the empty string).

In addition to the improved automaton construction procedure,
we also propose a refinement of the LR parsing paradigm which lends itself
to more intuitive automata.
Specifically, we examine the case of the {\em recursive-descent} (RD) parser:~intuitively, it
is easy to tell, at any point in time, what the parser is ``doing''. Namely, the stack trace
indicates that, e.g., it is ``parsing an expression'', and further up the stack,
we find that the expression is being parsed as the right-hand side of an assignment statement.
These qualitative statements are much harder to read off of a DFA state in the LR construction,
a fact which is largely responsible for the difficulties of debugging automatically generated parsers.

To remedy this, we introduce the concept of RD actions in an LR automaton.
Specifically, in addition to ``\texttt{Shift}'' and ``\texttt{Reduce} by production $\mc{P}$'',
our automata may call for the action ``\texttt{Recur} on top-level symbol $E$'' or
``\texttt{Return} from the current \texttt{Recur} call''.
The action on a state, e.g., $\mr{Stmt} \ra E := \cdot E$,
then becomes ``\texttt{Recur} on $E$'', rather than (as in standard LR)
having no action itself, but generating
$\eps$ prediction edges to all of the productions having $E$ as their left-hand side.
(We defer the full details to Section~\ref{sec:rd}.)
This simplification not only makes the resulting NFA states far more intuitive,
but also significantly reduces the proliferation of states, by reducing the need
for generated $\eps$ prediction edges.

\ \\
\noindent \textbf{Even canonical LR is not sufficiently general.}\ \ 
For full industrial programming languages, we often require functionality that significantly
exceeds the capabilities of canonical LR grammars.
For instance:
\begin{enumerate}
    \item Languages often require precedence specifications (e.g., $E \ra E + E$ with precedence $1$,
        and $E \ra E \times E$ with precedence $2$).
        While this requirement can, in principle, be satisfied by modifying the grammar to include new,
        explicit classes of nonterminals
        (in this case, ``terms''~$T$ and ``factors''~$F$),
        this operation significantly complicates the resulting grammar and renders the resulting syntax trees
        much more distant from their intuitive meanings.
        We would like an approach that does not require a transformation of the input grammar
        to satisfy the parser generator.
    \item Nonterminals may have semantic attributes---e.g., ``expression in a type context''
        versus ``expression in a value context''---that require separate treatment in order to be parsed
        unambiguously. While in principle this issue can be solved by creating a new nonterminal for each
        possible set of attributes, in practice this imposes untenable complexity on the grammar,
        as it requires a distinct nonterminal for every entry in the Cartesian product of attribute values
        (as well as again causing the resulting parse trees to be very far from their intuitive meanings).
    \item Nonterminals may have {\em syntactic} attributes that are necessary for unambiguous parsing.
        Classic examples include the ``\texttt{<>}'' template syntax in C++ and Rust,
        in which the expression enclosed in angle brackets cannot be an unguarded greater-than expression
        (lest the parser attempt to close the angle brackets upon scanning the ``\texttt{>}'' operator).
        Golang also contains many examples of this phenomenon, particularly when the specification
        requires that certain constructs be parsed greedily---e.g., ``\texttt{if * func () \{\}}''
        is a parse error, because \mbox{``\texttt{func () \{\}}''} must be parsed greedily as a lambda
        expression,
        leaving the if statement without a body (despite the fact that the alternative interpretation,
        treating ``\texttt{* func ()}'' as a starred expression, would otherwise be permitted).

        This type of requirement is notoriously difficult to capture in a declarative manner at the grammar level,
        a fact which is often invoked to support the conclusion that such languages can only be handled by
        hand-written recursive-descent parsers.
        In fact, however, we find that both cases can be represented by a form of (syntactic)
        attributes on nonterminals.
        In the case of angle-bracket templating, e.g.,
        the inner expression must have the attribute that it is not
        an ``\texttt{Expr.BinOp.Gt}'' production; while in the case of Golang's ``\texttt{if * func () \{\}}''
        example, we can require that the condition of an if statement have the attribute that
        it does not end with an unguarded function type ``\texttt{func ()}'' (a property which,
        in turn, is easy to specify at the level of productions of the grammar).
        We expand on this construction in more detail below.
    \item When represented in their most natural form, grammar rules often require intermediate reductions
        which lead to conflicts.
        For instance, a simplified version of Golang's struct field declaration takes the form
        $\mr{Field} \ra *?\,E\ |\ E\,E$ (where ``$*?$'' here indicates an optional star token).
        Although this construction is clearly easy to parse, and poses no challenge to a hand-written
        recursive-descent parser,
        a direct implementation of an LR parser will fail:~it will introduce
        an anonymous nonterminal for the ``$*?$'' expression (e.g., $X \ra *\ |\ \eps$),
        and, in cases in which the star is absent, will require the parser to decide whether to reduce
        by the production $X \ra \eps$ {\em immediately}, before seeing the token after
        the expression $E$ in order to determine which of the two $\mr{Field}$ cases applies.
        Intuitively, this issue does not seem fundamental to the parsing problem,
        but it nevertheless requires a general solution if we are to use LR parsing for full
        industrial languages.
    \item Finally, some common language constructs simply are not LR($k$) at a fundamental level.
        One example of such a construct is the {\em deferred branch}:~e.g.,
        $S \ra (\mr{Name} := \mr{Expr})\ |\ (\mr{Expr} = \mr{Expr})$,
        where we may have to make reduce decisions within $\mr{Name}$ and/or $\mr{Expr}$
        without yet knowing which of the two we are parsing.
        If the parser could proceed in parallel, it could shortly discard one of the two branches
        upon seeing a ``$:=$'' or ``$=$'' token, but standard LR cannot predict this until seeing
        a potentially unbounded number of tokens of the $\mr{Name}$ or $\mr{Expr}$ instance.
\end{enumerate}
In order to address these requirements, we introduce several new parser generation techniques,
which we now describe.

\ \\
\noindent \textbf{Technique: Nonterminal attributes.}
To address requirements 1-3, we first give a construction which permits nonterminals
to be accompanied by {\em attributes}, which can be used either syntactically or semantically.
In our proposal, attributes can be either binary-valued (as in most syntactic attributes),
or integer-valued (as in the case of operator precedence).
Relationships between attributes are then represented by {\em constraints} on grammar rules.
For example, in the case of operator precedence expressions, we might annotate rules as follows:
\[
    \begin{array}{rl}
        E \ra E + E  & \mr{rhs\_begin}[\mr{pr}] \geq 1 \vspace{0.5em} \\
                     & \mr{rhs\_end}[\mr{pr}] \geq 2 \vspace{0.5em} \\
                     & \mr{lhs}[\mr{pr}] \leq 1
    \end{array}
\]
\[
    \begin{array}{rl}
        E \ra E * E  & \mr{rhs\_begin}[\mr{pr}] \geq 2 \vspace{0.5em} \\
                     & \mr{rhs\_end}[\mr{pr}] \geq 3 \vspace{0.5em} \\
                     & \mr{lhs}[\mr{pr}] \leq 2
    \end{array}
\]
\[  \qquad  \]
\noindent while, for constraints such as in Golang's ``\texttt{if * func () \{\}}'' dilemma,
we might constrain the grammar rules as follows:
\[
\begin{array}{rl}
S \ra \mr{if}\ E_{[\tau]}\ \{ \ldots \} &  \mr{rhs\_tag}_{[\tau]}[\mr{mC}] \vspace{0.5em} \\
E \ra \mr{func}\ () &  \lnot \mr{lhs}[\mr{mC}] \vspace{0.5em} \\
E \ra E + E & \mr{lhs}[\mr{mC}] \leq \mr{rhs\_end}[\mr{mC}]
\end{array}
\]
(where the attribute ``mC'', intuitively, signifies an expression
that does not end with an unguarded function-type expression).\footnote{In
fact, the implementation of operator precedence in terms of constraints is slightly more subtle
than depicted here, because operators may have {\em left-precedence} and {\em right-precedence} attributes,
which require separate variables (e.g., \mr{prL} and \mr{prR}). For instance, the C++ expression \texttt{x.y()} requires that
``\texttt{.}'' bind tighter than ``\texttt{()}'', but this should not prevent \texttt{x().y}
from being a legal expression. A naive implementation, however, would stipulate that the
left-hand side of a dot should have precedence at least as high as the dot expression itself,
and thus would rule out the expression \texttt{x().y}.
With separate left-precedence and right-precedence, we avoid this issue:~a prefix operator only
constrains the right-precedence of its operands, while a postfix operator only constrains
the left-precedence.}

In order to implement these types of constraints, significant care is required
so that compilation of the LR automata remains tractable.
In particular, we restrict constraints to be {\em monotonic} in each attribute,
and {\em separable} across attributes; in other words,
we require every constraint to take one of the following forms:
\[
    \begin{array}{rl}
        & \mr{lhs}[\mr{attrA}] \leq C \vspace{0.5em} \\
        & \mr{rhs}[\mr{attrA}] \geq C \vspace{0.5em} \\
        & \mr{lhs}[\mr{attrA}] \leq \mr{rhs}[\mr{attrB}]
    \end{array}
\]
(keeping the convention that, for boolean-valued attributes, ``$\mr{attrA} \geq 1$'' corresponds to
the boolean constraint $\mr{attrA}$, and ``$\mr{attrA} \leq 0$'' corresponds to $\lnot \mr{attrA}$).

These restrictions rule out compound formulas involving multiple attributes,
as well as constraints that negate attributes;
thus, e.g., in order to express ``rhs[not attrA]'', it is necessary to introduce a separate attribute,
``rhs[attrNotA]''.
Despite the apparent severity of the restrictions, however,
we have found that virtually all syntactic and semantic attributes of interest
in industrial programming languages can be expressed in this fashion.
Moreover, the constraints remain declarative and compositional,\footnote{As
    opposed to constraints such as ``rule $\mc{P}$ may not be followed by a $\{$ token'',
    which are not compositional and may have unintended side-effects elsewhere in the grammar.
}
and thus can be implemented naturally as an extension of the LR automata.

\ \\
\noindent \textbf{Technique: CPS transformation.}
To address requirement 4 (intermediate reductions),
we propose a general grammar transformation, which we refer to as CPS
since its structure is reminiscent of the {\em continuation-passing style} technique
in functional programming.
At a high level, the purpose of the CPS transformation is to defer the decision
to reduce by a given production until the end of processing of the entire grammar rule
from which the production is derived.

Concretely, we illustrate the CPS transformation on the example above,
derived from Golang's grammar:
\[ \mr{Field}.\mr{Embedded} \ra *?\,E\ \]
\[ \mr{Field}.\mr{Standard} \ra \ E\,E \]
(where ``$*?$'' here indicates an optional star token).
Naively, when we flatten these syntax rules into a standard context-free grammar,
we would introduce an intermediate nonterminal (say, $X$),
and productions:
\[ \mr{Field} \ra X\,E \]
\[ X \ra * \]
\[ X \ra \eps \]
\[ \mr{Field} \ra E\,E \]
But this, of course, leads to the problem described above, namely that the parser must
decide whether to reduce by the production $X \ra \eps$ before seeing the input tokens
that follow the first expression $E$.

In the CPS-transformed version of such a grammar, we replace each intermediate nonterminal $X$
with a new nonterminal $X'$ which represents the concatenation of $X$ with the entire
remainder of the original rule in which $X$ appears.\footnote{
    We note that, for intermediate nonterminals generated from subexpressions of a given rule,
    there can be only one continuation of the nonterminal through the remainder of the rule,
    since the subexpression appears only once in the rule.
    For intermediate nonterminals generated as a result of iteration
    (e.g., a list of expressions),
    the translation is somewhat more complex, but follows the same principle.}
Thus, the CPS-transformed grammar would take the form:
\[ \mr{Field} \ra X' \]
\[ X' \ra * E \]
\[ X' \ra E \]
\[ \mr{Field} \ra E\,E \]
\noindent Formal details of the procedure are given in Section~\ref{sec:cps}.
Intuitively, we typically find that the CPS transformation does just what a human would do
when we inline rules manually to resolve conflicts in traditional LR parser generation.
By performing this transformation automatically, however, not only do we relieve the human
programmer from performing this inlining task, but we also enable the source grammar
(and hence the resulting syntax trees) to remain consistent with the programmer's
intention, as expressed in the original grammar prior to CPS transformation.

\ \\
\noindent \textbf{Technique: Parallel processing via XLR parsing.}
To address requirement 5,
we propose the following as a ``safety net'' for when LR, even in conjunction with the other
techniques of this work, fails to capture a language of interest.
Informally speaking, we define a $t$-parallel shift/reduce parser as one that may make nondeterministic
shift/reduce choices,
thereby spawning multiple parsing instances,
but which remains bounded to $t$ active instances throughout its execution on every input string.
An XLR($k, t$) grammar then is one that can be parsed by a $t$-parallel shift/reduce parser
with $k$ symbols of lookahead.

Our definition of XLR is similar to the established notion of Generalized LR (as in the ETL parsers),
but the important difference is that in XLR, we do not memoize the intermediate parsing results
for nonterminals on substrings.
This means that for grammars that fail to be XLR($k, t$) for bounded $t$,
an XLR parser could require exponential time
(potentially branching at every input symbol),
as opposed to the cubic time of Earley's and Leo's algorithms;
but, on the other hand, if $t$ is bounded, then the XLR running time
remains $O(tn)$, with a very small constant factor independent of $|\GG|$.
This allows XLR($k, t$) parsers with small values of~$t$ to remain competitive with hand-written
recursive-descent parsers, where the ETL algorithms would not be competitive.
Intuitively, XLR corresponds to recursive-descent parsing with a bounded amount of backtracking.

\ \\
\noindent \textbf{Technique: Conflict tracing.}
Finally, we introduce a technique which does not expand the class of grammars that our
parser generator recognizes, but which nevertheless greatly improves usability.
Specifically, we observe that one of the most common difficulties in using LR parser generators
in the wild is the opacity of conflicts.
When a conflict arises, it is often unclear which rules are responsible for it,
or what an example input might look like that would trigger the conflict.

To remedy this, we introduce a method of tracing conflicts in the LR automata,
which enables the reconstruction of a minimal pair of ``confusing inputs'' which triggers the conflict.
We implement conflict tracing by executing the standard subset construction
to convert the LR NFA to a DFA, and noting that, if conflicts occur,
then some DFA vertices will have conflicting accept actions.
For each conflicting DFA vertex $v$, we read off the labels that lead from the start vertex
to $v$. Then, we return to the original NFA, performing a search to identify paths
whose labels concatenate to the conflicting string in question.
The result is a collection of two (or more) paths in the NFA,
all traversed as a result of the same string of symbols,
which lead to distinct LR actions.
The completion of the ``confusing inputs'' can then be read off of the remainders
of each of the dotted productions along each path, since these are precisely what
is needed to complete the given productions and pop them off of the shift/reduce parsing stack.

Empirically, we have found that in translating real industrial grammars,
this conflict tracing method almost always pinpoints the precise cause of LR conflicts,
and the ``confusing inputs'' are indeed confusing to a human observer.
We also observe a surprising fact about the subset construction:~for real languages,
the DFA often contains {\em fewer} vertices than the original NFA,
despite the fact that in principle it could have exponential size.
We attribute this fact to the intuitive meaning of the LR DFA:~for real languages,
in order to be ``non-confusing'' to a human parser, the number of ``modes'' that the
parser can be in should be directly related to the size of the grammar.

\subsection{Related work}

The study of modern bottom-up (shift/reduce) parsing was pioneered by Knuth~\cite{Knu65},
who first defined the LR($k$) class of grammars.
Subsequent work by DeRemer~\cite{DeRemer} and others led to refinements including the
definition of SLR($k$) and LALR($k$) grammars,
which are less expressive than LR($k$) but which enable much more efficient parser generation,
and inspired the initial development of popular parser generation tools
such as Yacc~\cite{yacc} and Bison~\cite{bison}.
More recently,
a number of software tools, including Bison~\cite{bison}, Yacc++~\cite{yaccpp},
and gocc~\cite{gocc} have also implemented full LR parser generation.
As discussed above, our conclusion is that even LR($k$)---the most general of this family---is,
on its own, not sufficiently general to capture the grammars of real industrial languages
in a tractable manner.
However, we show that when combined with a number of features described in the
present work (e.g., optimized LR automata,
monotonic attributes, CPS transformation, and conflict tracing),
the fundamental LR paradigm does indeed provide the basis
for parser generation that is both general and practical.

Another line of work in bottom-up parsing is represented by the ETL family of algorithms~\cite{Earley, Tomita, Leo},
sometimes referred to as Generalized LR (GLR), and implemented
in software packages such as Marpa~\cite{Marpa} and Elkhound~\cite{Elkhound}.
These parsers process their input online, left-to-right, but instead of maintaining a single shift/reduce stack,
they maintain a dynamic programming table
capturing substrings of the input generated by particular dotted productions of the grammar.
The structure of this table means that unlike our parallel shift/reduce parsers, GLR parsers
maintain polynomial running time for all grammars.
On the other hand, however, for grammars which are in fact XLR($k, t$),
our parallel shift/reduce parsers maintain running time $O(tn)$
with a small constant factor independent of~$|\GG|$.

There has also been significant development of the top-down family of parser generators,
including tools such as ANTLR~\cite{ANTLR} which support a superset of LL($k$) grammars,
as well as their parallel counterparts, referred to as Generalized LL (GLL).
For the reasons described above, we believe that while top-down parsers are theoretically
interesting, the grammars they support are not sufficiently general to capture
full programming languages.\footnote{We
    note that languages such as Python,
    which have BNF grammars that are used in top-down parsing style,
    actually depend on procedural features (e.g., PEGs)
    rather than being purely declarative grammars.}
Notably, they are significantly more restrictive than LR($k$),
and we argue that even the latter is not sufficiently general in its standard formulation.

A completely different approach to parsing is taken by the {\em procedural} family
of parser generators,
which, instead of encoding a CFG directly in a declarative fashion,
use the ordering of rules to dictate matching priority and resolve ambiguities.
In this respect, they can be viewed as an optimized toolkit for constructing
recursive-descent parsers more efficiently than writing code by hand---the
parser definition describes which productions to try in which order,
just as one does in backtracking recursive-descent.
Examples of these parsers are the Parser Expression Grammars (PEGs) of~\cite{PEG},
the corresponding ``packrat'' family of parsers~\cite{packrat},
and a long line of work on parser combinators~(e.g., the Parsec library in Haskell \cite{parsec}).
This structured approach is often superior to hand-written recursive descent
in implementation cost and maintainability, though it still suffers from many of the same
disadvantages:~namely, there is no declarative context-free specification of the grammar,
and as a result, it is often difficult to debug and verify that the grammar captures
the programmer's intent.
In this work, we confine our attention to fully-declarative rather than procedural parsing.

We also mention two existing algorithms for constructing optimized LR automata.
First, there is a classic algorithm of Pager~\cite{Pager}.
Like our construction, it yields automata that are much smaller
than standard LR,
but the mechanism is different:~rather than constructing LR($0$) DFA states
and inferring $k$-follow set refinements necessary to resolve conflicts, it constructs full
canonical LR($1$) states,
then examines them to merge those that can be merged without creating new conflicts.
Because of this, while the final resulting DFA may be quite small, the intermediate computation
can be significantly more expensive due to the processing of the entire LR($1$) automaton.
Second, there is the IELR algorithm of Denny and Malloy~\cite{DM10}.
IELR is somewhat similar to the backward phase of our algorithm,
in that it identifies conflicts at reduction vertices and propagates them backwards
to split earlier vertices.
However, IELR operates on the DFA rather than the NFA,
and first requires the formation of the LALR($1$) automaton
in order to begin splitting states.

Finally, we mention one other result that refines the implementation of LR($k$) parsing
in a similar direction to the present work.
The tree automata of Adams and Might~\cite{AM17} are similar to our implementation of
monotonic attributes; they differ in that they impose left-to-right sequencing
on the attribute constraints
(whereas attributes can refer to arbitrary constituents of a grammar rule),
and require productions to be duplicated (whereas attributes
are boolean or integer values layered independently on top of the standard context-free grammar rules).

\section{Definitions}
\label{sec:defs}

Fix a context-free grammar $\GG$
over nonterminal alphabet $\Lambda$,
terminal alphabet $\Sigma$, and
extended terminal alphabet $\Gamma = \Sigma \cup \{\srt\}$,
where $\srt$ is the standard right-end sentinel marker.
We assume the start symbol $S$ of $\GG$
does not appear in the right-hand side of any production,
and appears in the left-hand side of exactly one production
(this is without loss of generality, since we may always introduce
a new start symbol $S' \ra S$).
The empty string, as usual, is denoted by $\eps$.
We denote concatenation of strings $\alpha,\beta$
as $\alpha\beta$,
and we generalize this in the standard way
to concatenations that include sets of strings.

\begin{definition}[$k$-Prefix Set]\normalfont
\label{def:kpref}
Fix an integer $k \geq 0$,
and a set of strings $T \subseteq \Gamma^*$.
We define the $k$-prefix set, $\fst_k(T)$,
as the set of all strings $x_1\cdots x_{\min(k,r)}$
for $x_1\cdots x_r \in T$.
\end{definition}

\begin{definition}[Generated Set]\normalfont
\label{def:gen}
Fix a set of strings $T \subseteq (\Lambda \cup \Gamma)^*$.
We define the generated set, $\fstgen(T)$,
as the set of all strings $y$
such that $x \genstar y$ for some $x \in T$.
\end{definition}

\begin{definition}[Partition]\normalfont
Sets $T_1,\ldots,T_m \subseteq S$ are a partition of $S$ if
$T_i \cap T_j = \emptyset$ for all $i \neq j$,
and $\bigcup_i T_i = S$.
\end{definition}

\begin{definition}[Partition Refinement]\normalfont
Suppose $T$ and $U$ are partitions of a set $S$.
Then we define the refinement of $T$ by $U$
to be the partition $\{ T_i \cap U_j : T_i \in T, U_j \in U \}$.
(Note that this is a partition by construction,
since if $(T_i \cap U_j) \cap (T_{i'} \cap U_{j'}) \neq \emptyset$,
then $i = i'$ and $j = j'$.)
\end{definition}

\begin{definition}[$k$-Follow Set Partition]
    \normalfont
    \label{def:klook}
Fix an integer $k \geq 0$.
We define a $k$-follow set partition~$L$ of the grammar $\GG$
to comprise the following, for each dotted production $\hat \Pi$ of $\GG$:
\begin{itemize}
    \item A universe $\mc{U}(\hat \Pi) \subseteq \Gamma^k$ of possible $k$-follow strings.
    \item A partition $L(\hat \Pi)$ of $\mc{U}(\hat \Pi)$.
\end{itemize}
\end{definition}

Without loss of generality, we will always take $\mc{U}(\hat \Pi)$ to be
the set of strings in $\Gamma^k$ that can ever follow the base production $\Pi$
in derivations of the grammar $\GG$;
this can be computed efficiently by a flood-fill
algorithm, as in standard SLR parsing.

In order to define the LR($k$) automata, we also recount the definition
of a nondeterministic finite automaton (NFA).
We note that our definition differs somewhat from standard definitions,
since
we allow multiple actions on a given vertex,
rather than at most one,
and we designate particular pairs of actions as ``conflicting''.
This aspect of the definition will be important in defining lookahead-dependent
LR actions, since we will need to specify actions such as ``$a$ on lookahead $\lambda$,
$a'$ on lookahead $\mu$''
(which conflict only if $a \neq a'$ and $\lambda = \mu$).

\begin{definition}[Nondeterministic Finite Automaton]\normalfont
    \label{def:nfa}
Fix sets $V$ (the vertices), $A$ (the alphabet), and $T$ (the accepting actions),
and assume a relation ``$\mr{conf}$'' on $T \times T$,
intuitively signifying when two actions conflict with one another.
We extend the relation $\mr{conf}$ to sets of actions
by specifying that for sets $U,V \subseteq T$,
we have $\mr{conf}(U,V)$ if for some $U_i \in U$ and $V_j \in V$, we have $\mr{conf}(U_i,V_j)$;
and, for a single set $U \subseteq T$,
we have $\mr{conf}(U)$ if $\mr{conf}(U,U)$.
Then an NFA $N$ consists of the following.
\begin{itemize}
    \item A designated vertex $v_s \in V$ (the starting vertex).
    \item For each vertex $v \in V$, a set of accepting actions, $T_v \subseteq T$.
    \item A set of edges $v \overset{\lambda}{\ra} w$, with $v,w \in V$, and 
        some label $\lambda \in A \cup \{\eps\}$.
\end{itemize}
\end{definition}

Intuitively, we will generally take $A = \Lambda \cup \Gamma$,
$V$ the set of LR vertices,
and $T$ the set of LR actions associated to each possible $k$-token
lookahead in the input string.

\begin{definition}[NFA Path (Sequence)]\normalfont
    \label{def:nfapathseq}
    A path in an NFA $N$ on input sequence $\tau = \tau_1 \cdots \tau_r$,
    where each $\tau_i \in A \cup \{\eps\}$,
    consists of a sequence of vertices $v_1,\ldots,v_{r+1}$
    such that $v_1 \in S$ is the starting vertex and,
    for each $i \in [r]$,
    there exists an edge from $v_i$ to $v_{i+1}$ on the corresponding label $\tau_i$.
    The accepting action set for such a path
    is defined to be the accepting action set $T_{v_{r+1}} \subseteq T$ of the final vertex $v_{r+1}$.
\end{definition}

\begin{definition}[NFA Path (String)]\normalfont
    \label{def:nfapathstr}
    We extend Definition~\ref{def:nfapathseq} to strings $\zeta \in A^*$
    as follows.
    $P$ is a path in an NFA~$N$ on a string $\zeta$
    if there exists a sequence $\tau = \tau_1 \cdots \tau_r$
    such that $P$ is a path in $N$ on the sequence $\tau$,
    and $\tau_1 \cdots \tau_r = \zeta$.
\end{definition}

\begin{definition}[NFA Conflict]\normalfont
    A conflict in an NFA $N$ consists of two NFA paths
    on the same input string $\zeta$
    whose accepting action sets conflict.
\end{definition}

\begin{definition}[Deterministic Finite Automaton]\normalfont
    Fix sets $V$ (the vertices), $A$ (the alphabet), and $T$ (the accepting actions),
    and a relation ``$\mr{conf}$'' on $T \times T$ as in Definition~\ref{def:nfa}.
    Then a DFA $D$ consists of the following.
    \begin{itemize}
        \item A starting vertex $S_0 \in V$.
        \item For each vertex $v \in V$, a set of accepting actions, $T_v \subseteq T$.
        \item A set of edges $v \overset{\lambda}{\ra} w$, with $v,w \in V$ and 
            $\lambda \in A$,
            such that for each vertex $v$
            and each label $\lambda \in A$,
            there exists exactly one vertex $w$ with an edge $v \overset{\lambda}{\ra} w$.
    \end{itemize}
\end{definition}

We extend Definitions~\ref{def:nfapathseq} and \ref{def:nfapathstr}
to DFA paths in the natural way.

\begin{definition}[DFA Conflict]\normalfont
    A conflict in a DFA~$D$ consists of a path from the starting vertex
    to a vertex whose accepting action set conflicts.
\end{definition}

We now present a definition of the LR($k$)~NFA,
which generalizes both the SLR and canonical LR automata.

\begin{construction}[LR($k$) NFA]\normalfont
\label{cons:lrknfa}
Fix an integer $k \geq 0$,
and a $k$-follow set partition $L$ of the grammar $\GG$ (Definition~\ref{def:klook}).
We define the LR($k$) NFA of $\GG$ over $L$ as follows
(assuming the convention that accepting actions may have associated lookaheads,
and two distinct actions conflict if they have the same lookahead).
\begin{itemize}
    \item The vertices consist of all pairs $(\hat \Pi,\,\bar \lambda)$
        where $\hat \Pi$ is a dotted production of the grammar and
        $\bar \lambda \in L(\hat \Pi)$.
    \item The starting vertex is $(S \ra \cdot\,\eta,\,\bar \lambda)$,
        where $S \ra \eta$ is the production whose left-hand side is the start symbol $S$,
        and $\bar \lambda$ is the set in the partition such that $\srt^k \in \bar\lambda$.
    \item The edges are determined as follows:
    \begin{itemize}
        \item Prediction edges: for vertices $v = (X \ra \alpha \cdot Y \beta,\,\bar \lambda)$
            and $w = (Y \ra \cdot\,\gamma,\, \bar \mu)$,
            there is an $\eps$-edge from $v$ to $w$
            whenever $\bar \mu \cap \fst_k(\fstgen(\beta \bar \lambda)) \neq \emptyset$.
        \item Step edges: for vertices $v = (X \ra \alpha \cdot \tau \beta,\,\bar \lambda)$
            and $w = (X \ra \alpha \,\tau \cdot \beta,\,\bar \lambda)$,
            there is an edge from $v$ to $w$ with label $\tau$
            (where $\tau$ is a single symbol, either terminal or nonterminal).
    \end{itemize}
    \item The vertex accept actions are determined as follows:
    \begin{itemize}
        \item For vertices $(X \ra \alpha \, \cdot\,,\,\bar \lambda)$,
            for every $\lambda \in \bar \lambda$,
            the associated action on lookahead $\lambda$ is ``Reduce $X \ra \alpha$''.
        \item For vertices $(X \ra \alpha \cdot \sigma \beta,\, \bar \lambda)$,
            where $\sigma \in \Sigma$ is a terminal symbol,
            for every $\mu \in \fst_k(\fstgen(\sigma \beta \bar \lambda))$,
            the associated action on lookahead $\mu$ is ``Shift''.
    \end{itemize}
\end{itemize}
\end{construction}

\begin{construction}[LR($k$) DFA]\normalfont
\label{cons:lrkdfa}
For integer $k \geq 0$,
and a $k$-follow set partition $L$ of $\GG$,
we define the LR($k$) DFA of $\GG$ over $L$
to be the automaton obtained by running the standard subset construction
on the LR($k$) NFA (Construction~\ref{cons:lrknfa}),
taking unions over the resulting accept action sets at each vertex.
\end{construction}

For $k = 0$, there is only one possible $k$-follow set partition
(namely, $L(\hat \Pi) = \{\{\eps\}\}$
for every dotted production $\hat \Pi$),
and thus we may refer to the LR($0$)~NFA/DFA without ambiguity.
For $k > 0$, the LR($k$) automata as we have defined them depend on the particular choice
of partition $L$.
If we take $L$ to be the coarsest partition possible,
then we recover the definition of the SLR automaton:
\begin{definition}[SLR($k$) NFA]\normalfont
    The {\em SLR($k$) NFA} is the LR($k$) NFA given by Construction~\ref{cons:lrknfa}
    applied to the partition
    $L^*(\hat \Pi) = \{ \,\mc{U}(\hat \Pi) \}$ for every dotted production $\hat \Pi$.
\end{definition}
\noindent On the other hand, if we take $L$ to be the most refined partition possible,
then we recover the standard definition of the canonical LR($k$) automaton.
\begin{definition}[Canonical LR($k$) NFA]\normalfont
    The {\em canonical LR($k$) NFA} is the LR($k$) NFA given by Construction~\ref{cons:lrknfa}
    applied to the partition
    $L^*(\hat \Pi) = \{ \{ \tau \} : \tau \in \mc{U}(\hat \Pi) \}$ for every dotted production~$\hat \Pi$.
\end{definition}

We also note that if we consider the LR($k$) NFA over some partition $L$ coarser than $L^*$,
we find that it is isomorphic to the NFA produced by taking the canonical LR($k$) automaton
and grouping vertices by the equivalence relation defined by the partition $L$.
This property will be important when we consider the task of optimizing LR($k$) automata,
as we will ultimately aim to emulate the result of grouping the vertices of the canonical
LR($k$) automaton by some equivalence relation,
but we will avoid explicitly constructing the canonical automaton,
instead invoking Construction~\ref{cons:lrknfa} on an appropriate partition.

Finally, for completeness, we also recount the standard stack-based
LR parsing algorithm on an LR($k$) DFA.
\begin{construction}[Stack-Based LR Parsing Algorithm]\normalfont
    \label{cons:stacklrparse}
    Fix a grammar $\GG$ and an integer $k$
    and let $D$ be a conflict-free LR($k$)~DFA (Construction~\ref{cons:lrkdfa})
    for $\GG$.
    Let $S_0$ denote the starting state of~$D$,
    and fix an input string $x = x_1\ldots x_n$.
    The stack-based LR parsing algorithm on $x$ then operates as follows.
    \begin{enumerate}
        \item Initialize the state stack with a single element, $S_0$,
            and initialize the syntax stack to be empty.
            Initialize the cursor position $i = 1$.
        \item Repeat:
        \begin{enumerate}
            \item If the state stack is empty, halt with failure.
            \item Let $v$ be the vertex on top of the state stack.
                Examine the current lookahead $\lambda$, i.e., the first $k$ symbols of
                $x_i x_{i+1} \ldots x_n \srt^k$.
                If $v$ has no action on $\lambda$, then halt with failure.
                Otherwise, act according to the action on $\lambda$:
                \begin{itemize}
                    \item If the action is ``Shift'',
                        then set the current buffer symbol to $x_i$,
                        push $x_i$ onto the syntax stack,
                        and increment $i$
                        (unless $i = n+1$, in which case halt with failure).
                    \item If the action is ``Reduce $\Pi = X \ra \alpha_1 \ldots \alpha_s$''
                        then pop $s$ elements off of the syntax stack,
                        construct a syntax element labeled ``$\Pi$'' with those $s$ arguments
                        as children,
                        and push the resulting element back onto the syntax stack.
                        Then, pop $s$ elements off of the state stack,
                        and set the current buffer symbol to $X$.
                \end{itemize}
            \item Let $\alpha$ denote the current buffer symbol.
                \begin{itemize}
                    \item If $\alpha = S$, then if $i = n+1$, halt with success
                        and return the top (i.e., only) element of the syntax stack;
                        otherwise, halt with failure.
                    \item If $\alpha \neq S$,
                        then let $v'$ be the vertex for which there is an edge $v \overset{\alpha}{\ra} v'$
                        (there is exactly one such $v'$, since $D$ is a DFA).
                        Push $v'$ onto the state stack.
                \end{itemize}
        \end{enumerate}
    \end{enumerate}
\end{construction}

\section{Results}

We now present the main theoretical contributions of this work, namely
several new developments of the LR parsing paradigm.
For simplicity of presentation, we introduce each development separately
as an extension of canonical LR,
though in our software implementation they are enabled simultaneously,
along with a number of additional syntactic conveniences.

\subsection{Grammar flattening}
\label{sec:grmflat}

We begin by describing the basic mechanism by which we translate a full BNF-style grammar,
which contains various syntactic conveniences, to a standard CFG.
This transformation will take, e.g., rules such as the following
for a multidimensional array index expression:\footnote{To avoid confusion,
note that \texttt{[} and \texttt{]} are terminal symbols, while parentheses $(\ldots)$
group symbols at the BNF (meta) level.}
\[ E \ra E \ \texttt{[}\ E \  (,\,E)^* \  ,\hspace*{-0.2em}?\ \texttt{]}  \]
generate intermediate nonterminals $X, Y$ for the nontrivial subexpressions of the rule,
and produce flat CFG productions such as the following:
\[ E \ra E \ \texttt{[}\ E\ X\ Y \ \texttt{]} \]
\vspace*{-1.65em}
\[ X \ra \ ,\hspace*{-0.2em}\ E\ X \]
\[ X \ra \eps \]
\[ Y \ra \ , \]
\[ Y \ra \eps \]
Other constructs are handled in a similar manner.
We give a full description in Construction~\ref{cons:grmflat}.

\begin{definition}[BNF Grammar]\label{def:bnf}\normalfont
    A BNF grammar over nonterminal symbols $\Lambda$
    and terminal symbols $\Sigma$
    consists of a sequence of {\em rules}
    of the form $X \ra e$,
    where $X \in \Lambda$ is a nonterminal,
    and $e$ is an expression defined inductively as follows:
    \begin{itemize}
        \item $e = \sigma$, where $\sigma \in \Lambda \cup \Sigma$ is a single symbol.
        \item $e = e_1\ e_2\ \ldots\ e_n$, a concatenation of expressions.
        \item $e = e_1\ |\ e_2\ |\ \ldots\ |\ e_n$, an alternation of expressions.
        \item $e = e_1?$, an optional expression.
        \item $e = e_1^*$, a repeated expression.
    \end{itemize}
\end{definition}

\begin{construction}[Grammar Flattening]\label{cons:grmflat}\normalfont
Grammar flattening is an inductive definition, specified by the procedure
$\mr{Flatten}(X, e)$,
where $X$ is a target symbol and $e$ is a BNF expression
as given by Definition~\ref{def:bnf}.
To flatten an entire grammar, invoke $\mr{Flatten}(X, e)$ for each BNF rule $X \ra e$.
The inductive cases of the procedure $\mr{Flatten}(X, e)$ as follows:
\begin{itemize}
    \item $e = \alpha$ where $\alpha$ is a single symbol (either terminal or nonterminal).
        Emit the rule $X \ra \alpha$.
    \item $e = e_1\ e_2\ \ldots\ e_n$, a concatenation.
        For each $i$, if $e_i$ is a single symbol, let $\alpha_i = e_i$;
        otherwise, let $\alpha_i$ be a fresh nonterminal, and invoke $\mr{Flatten}(\alpha_i, e_i)$.
        Emit the rule $X \ra \alpha_1\ \alpha_2\ \ldots\ \alpha_n$.
    \item $e = e_1\ |\ e_2\ |\ \ldots\ |\ e_n$, an alternation.
        For each $i$, invoke $\mr{Flatten}(X, e_i)$.
    \item $e = e_1?$, an optional expression.
        Invoke $\mr{Flatten}(X, e_1\ |\ \eps)$.
    \item $e = e_1^*$, a repeated expression.
        Let $\alpha$ be a fresh nonterminal. Invoke $\mr{Flatten}(\alpha, e_1)$.
        Emit rules $X \ra \alpha\,X$ and $X \ra \eps$.
\end{itemize}
Additional convenience constructs, such as $e = e_1^+$ (a repeated expression with at least one instance),
and $e = \mr{List}[e_1 :: e_2]$ (a list of expression $e_1$ with delimiter $e_2$),
may be handled analogously to the repeated expression case ($e = e_1^*$) above.
Our software implementation includes these and other syntactic conveniences.
\end{construction}

\subsection{CPS transformation}
\label{sec:cps}

Once a grammar is flattened, we perform an additional transformation
which we call CPS, as it is reminiscent of the {\em continuation-passing style}
technique in functional programming.
At a high level, the purpose of the CPS transformation is to eliminate conflicts
by deferring reductions
of intermediate productions until the latest possible moment,
namely the end of processing the entire BNF rule.
This is accomplished by replacing each intermediate nonterminal (i.e., those generated during flattening)
with a CPS version, which corresponds to the content of that nonterminal concatenated
with the entire remainder of the rule in which it appears.
Following our example from Section~\ref{sec:grmflat},
we would produce the following CPS-transformed rules:
\[ E \ra E\ \texttt{[}\ E\ X' \]
\vspace*{-1.62em}
\[ X' \ra\ ,\hspace*{-0.2em}\ E\ X' \]
\[ X' \ra Y' \]
\[ Y' \ra\ ,\hspace*{-0.2em}\ \texttt{]} \]
\[ Y' \ra\ \texttt{]} \]
\vspace*{-1em} \\ \ 
Note that the tail of the production, $Y'$, is ``absorbed'' into the base case
of the recursive symbol,~$X'$, rather than appearing in the inductive case.

Additionally, a key property of the CPS transformation is that the output remains linear
in the size of the full input BNF rule.
This may be somewhat counterintuitive, but can best be illustrated with an example such as
the following BNF rule:
\[ S \ra (A\ |\ B)\ (C\ |\ D)\ (E\ |\ F) \]
The flattening procedure transforms such a rule into the following productions:
\[ S \ra X_0\ X_1\ X_2 \]
\vspace*{-1.62em}
\[ X_0 \ra A \]
\[ X_0 \ra B \]
\[ X_1 \ra C \]
\[ X_1 \ra D \]
\[ X_2 \ra E \]
\[ X_2 \ra F \]
\vspace*{-1em} \\ \ 
The CPS transformation then yields the following grammar:
\[ S \ra Y_0 \]
\vspace*{-1.62em}
\[ Y_0 \ra A\ Y_1 \]
\[ Y_0 \ra B\ Y_1 \]
\[ Y_1 \ra C\ Y_2 \]
\[ Y_1 \ra D\ Y_2 \]
\[ Y_2 \ra E \]
\[ Y_2 \ra F \]
\vspace*{-1em} \\ \ 
We note that although each original symbol $X_i$ has been transformed to a CPS symbol $Y_i$
which subsumes the remainder of its BNF rule,
these symbols $Y_i$ are reused in the tails of the productions, and thus avoid
duplication of the tail contents
(as would occur in a naive translation: $S \ra A\ C\,E$, $S \ra A\ C\,F$, etc.)

We now give the full definition of the transformation.

\begin{construction}[CPS Transformation]\label{cons:cps}\normalfont
We define a nonterminal $X \in \Lambda$ to be {\em CPS-eligible}
if it was generated as a fresh symbol during the BNF flattening procedure
(Construction~\ref{cons:grmflat}),
as opposed to a top-level symbol of the BNF grammar.
We further fix a subset of the CPS-eligible nonterminals
that are specified to be {\em CPS-triggering}.
(In a concrete implementation, these may be all CPS-eligible symbols;
or, in an optimized implementation, only the subset of such symbols
which might otherwise trigger an SLR reduction conflict
if not CPS-expanded.)
For nonterminals $X \in \Lambda$, let $\hat X$ be a fresh symbol
(intuitively, the CPS-transformed version of $X$).

We now define mutually inductive procedures
$\mr{CPSProd}(X \ra \alpha_1 \ldots \alpha_i \cdot \alpha_{i+1} \ldots \alpha_r,\,\tau)$,
where $X \ra \alpha_1 \ldots \alpha_i \cdot \alpha_{i+1} \ldots \alpha_r$
is a dotted production of the flattened grammar,
and $\tau$ is a sequence of symbols (intuitively, the tail context);
and $\mr{CPSSym}(X,\,\tau)$ where $X$ is a symbol.
To define the procedure $\mr{CPSProd}(X \ra \alpha_1 \ldots \alpha_i \cdot \alpha_{i+1} \ldots \alpha_r,\,\tau)$,
we consider the following cases:
\begin{itemize}
    \item The dot is at the beginning of the production (i.e., $i = 0$).
        In this case, add the production $\hat X \ra \tau$ to the CPS grammar.
    \item The dot is not at the beginning of the production, and the symbol
        $\alpha_i$ preceding the dot is not CPS-triggering.
        In this case, invoke
        $\mr{CPSProd}(X \ra \alpha_1 \ldots \alpha_{i-1} \cdot \alpha_i \ldots \alpha_r,\,\hat\alpha_i\tau)$.
    \item The dot is not at the beginning of the production, and the symbol
        $\alpha_i$ preceding the dot is CPS-triggering.
        We consider the following subcases:
        \begin{itemize}
            \item The dotted production $X \ra \alpha_1\ldots\alpha_i \cdot \alpha_{i+1} \ldots \alpha_r$
                is of the form $\alpha \ra \gamma \alpha\, \cdot$,
                where $\alpha$ is CPS-triggering (i.e., the production was generated
                as a result of flattening a repeated expression).
                In this case, invoke $\mr{CPSProd}(\alpha \ra \gamma \cdot \alpha,\,\hat \alpha)$.
            \item The dotted production is not of the indicated form.
                In this case, invoke $\mr{CPSSym}(\alpha_i,\,\tau)$,
                and finally invoke
                $\mr{CPSProd}(X \ra \alpha_1 \ldots \alpha_{i-1} \cdot \alpha_i \ldots \alpha_r,\,\hat \alpha_i)$.
        \end{itemize}
\end{itemize}
To define the procedure $\mr{CPSSym}(X,\,\tau)$,
we proceed as follows:
\begin{itemize}
    \item For each production $X \ra \eta$ of the original flattened grammar,
        invoke $\mr{CPSProd}(X \ra \eta\,\cdot,\,\tau)$.
\end{itemize}
Finally, to CPS-transform an entire flattened grammar,
we simply
invoke $\mr{CPSSym}(X,\,\eps)$
for each non-CPS-triggering nonterminal $X$.
\end{construction}

A key property of the CPS transformation is that it leaves the language
generated by the grammar unchanged. We can show this by defining an appropriate
relation between strings generated by symbols $X$ and those generated by
their CPS counterparts $\hat X$.
\begin{definition}[CPS Tail Context]\normalfont
For a symbol $X$ of the original flattened grammar,
we define the {\em CPS tail context} $\tau_X$ as follows:
\begin{itemize}
    \item If $X$ is a terminal symbol, define $\tau_X = \eps$.
    \item If $X$ is a nonterminal symbol,
        define $\tau_X$ to be the string such that $\mr{CPSSym}(X,\tau_X)$
        is invoked in the CPS transformation of the grammar.
\end{itemize}
For a dotted production $\Pi = X \ra \alpha_1\ldots \alpha_i \cdot \alpha_{i+1} \ldots \alpha_r$,
we define the tail context analogously,
to be the string $\tau_\Pi$ such that $\mr{CPSProd}(\Pi,\tau_\Pi)$
is invoked in the CPS transformation.

It remains to show that this criterion is well-defined, i.e., that for each nonterminal $X$,
the procedure $\mr{CPSSym}(X,\cdot)$ is invoked exactly once.
This is evident for non-CPS-triggering nonterminals $X$;
for these nonterminals, the tail context $\tau_X$ is simply $\eps$,
since $\mr{CPSSym}$ can only be invoked on $X$ at top level.
In the general case, we first note that by construction,
$\mr{CPSSym}$ only makes recursive calls on the productions $X \ra \eta$
for the input symbol $X$,
and $\mr{CPSProd}$ makes a single call to $\mr{CPSSym}$
for each occurrence of a CPS-triggering symbol in the right-hand side
of the production (excluding those arising from 
repeated expressions, $\alpha \ra \gamma \alpha\,\cdot$).
Since every CPS-triggering symbol is CPS-eligible,
and every CPS-eligible symbol is reachable from a non-CPS-triggering symbol
(and appears exactly once in the right-hand side of a production, excluding its occurrence in
repeated expressions $\alpha \ra \gamma \alpha\,\cdot$),
we conclude that $\mr{CPSSym}$ is invoked exactly once on every nonterminal.
\end{definition}
\begin{lemma}[CPS Relation, Forward]\normalfont
    \label{lem:cpsfwd}
The following properties hold:
\vspace*{-0.2em}
\begin{enumerate}
\item For every symbol $X$ of the flattened grammar,
for every terminal string $\lambda \in \Sigma^*$,
if $X \genstar \lambda$ in $n$ steps,
then $\hat X \genstar \lambda \tau_X$ in $n$ steps.
\item For every dotted production
$\Pi = X \ra \alpha_1\ldots \alpha_{i-1} \cdot \alpha_i \ldots \alpha_r$,
for every terminal string $\lambda \in \Sigma^*$,
if $\alpha_i \ldots \alpha_r \genstar \lambda$ in $n$ steps,
then $\tau_\Pi \genstar \lambda \tau_X$ in $n$ steps.
\end{enumerate}
\end{lemma}
\begin{proof}
    By induction on the pair $(n, |\lambda|)$ in lexicographic order.
    To show property $1$, first note that if $n = 0$,
    then $X$ is a terminal symbol, and the claim trivially holds.
    So, suppose $n > 0$,
    and $X \ra \alpha_1 \ldots \alpha_r \genstar \mu$ in $n$ steps.
    Then invoking property $2$ on the dotted production $X \ra \cdot\,\alpha_1 \ldots \alpha_r$,
    we find that $\tau_\Pi \genstar \mu \tau_X$ in $(n-1)$ steps.
    Since the $\mr{CPSProd}$ procedure with the dot at the beginning
    outputs precisely the rule $\hat X \ra \tau_\Pi$,
    we find that $\hat X \genstar \lambda \tau_X$ in $n$ steps, as desired.

    To show property $2$, first consider the case in which the
    dot is at the end. Then $\tau_\Pi = \tau_X$ by definition,
    and since $\alpha_i \ldots \alpha_r$ is the empty string,
    $\lambda$ is the empty string,
    and hence $\tau_\Pi = \lambda \tau_X \genstar \lambda \tau_X$ in $0$ steps, as desired.
    Now consider the case in which the dot is not at the end.
    Let $\Pi' = X \ra \alpha_1 \ldots \alpha_i \cdot \alpha_{i+1} \ldots \alpha_r$.
    We consider the following cases:
    \begin{itemize}
        \item $\alpha_{i}$ is not CPS-triggering.
            Then $\tau_\Pi = \hat\alpha_i \tau_{\Pi'}$.
            Suppose $\lambda = \mu \nu$, where $\alpha_i \genstar \mu$ in $n_\mu$ steps
            and $\alpha_{i+1}\ldots \alpha_r \genstar \nu$ in $n_\nu$ steps,
            with $n = n_\mu + n_\nu$.
            Now clearly $n_\mu, n_\nu \leq n$, and furthermore:
            \begin{itemize}
                \item If $\alpha_i$ is a terminal symbol, then $|\mu| = 1$, hence $|\nu| < |\lambda|$.
                \item If $\alpha_i$ is a nonterminal symbol, then $n_\mu > 0$, hence $n_\nu < n$.
            \end{itemize}
            We conclude inductively by property $1$ that
            $\hat \alpha_i \genstar \mu \tau_{\alpha_i} = \mu \eps = \mu$ in $n_\mu$ steps,
            and by property $2$ that $\tau_{\Pi'} \genstar \nu \tau_X$ in $n_\nu$ steps.
            Hence $\tau_\Pi = \hat \alpha_i \tau_{\Pi'} \genstar \mu \nu \tau_X = \lambda \tau_X$
            in $n_\mu + n_\nu = n$ steps,
            as desired.
        \item $\alpha_i$ is CPS-triggering, and $\Pi$ is of the form $\alpha_i \ra \gamma \cdot \alpha_i$.
            Then $\tau_\Pi = \hat \alpha_i$,
            and we conclude inductively by property $1$ that
            $\tau_\Pi = \hat \alpha_i \genstar \lambda \tau_{\alpha_i}$ in $n$ steps.
        \item $\alpha_i$ is CPS-triggering, and $\Pi$ is not of the indicated form.
            Then $\tau_{\alpha_i} = \tau_{\Pi'}$,
            and $\tau_\Pi = \hat \alpha_i$.
            Suppose $\lambda = \mu \nu$, where $\alpha_i \genstar \mu$ in $n_\mu$ steps
            and $\alpha_{i+1}\ldots \alpha_r \genstar \nu$ in $n_\nu$ steps,
            with $n = n_\mu + n_\nu$.
            Now clearly $n_\mu, n_\nu \leq n$, and furthermore $n_\mu > 0$ (since $\alpha_i$ is a nonterminal),
            hence $n_\nu < n$.
            We conclude inductively by property $1$ that 
            $\hat \alpha_i \genstar \mu \tau_{\alpha_i} = \mu \tau_{\Pi'}$ in $n_\mu$ steps,
            and by property $2$ that $\tau_{\Pi'} \genstar \nu \tau_X$ in $n_\nu$ steps.
            Hence $\tau_\Pi = \hat \alpha_i \genstar \mu \tau_{\Pi'} \genstar \mu \nu \tau_X = \lambda \tau_X$
            in $n_\mu + n_\nu = n$ steps,
            as desired. \qedhere
    \end{itemize}
\end{proof}

\begin{lemma}[CPS Relation, Reverse]\normalfont
    \label{lem:cpsrev}
    The following properties hold:
    \vspace*{-0.2em}
    \begin{enumerate}
    \item For every symbol $X$ of the flattened grammar,
        if $\hat X \genstar \lambda$ in $n$ steps,
        then $X \tau_X \genstar \lambda$ in $n$ steps.
    \item For every dotted production
        $\Pi = X \ra \alpha_1\ldots \alpha_{i-1} \cdot \alpha_i \ldots \alpha_r$,
        if $\tau_\Pi \genstar \lambda$ in $n$ steps,
        then $\alpha_i \ldots \alpha_r \tau_X \genstar \lambda$ in $n$ steps.
    \end{enumerate}
\end{lemma}
\begin{proof}
        By induction on the pair $(n, |\lambda|)$ in lexicographic order.
        To show property~$1$, first note that if $n = 0$,
        then $\hat X$ is a terminal symbol, and the claim trivially holds.
        So, suppose $n > 0$,
        and $\hat X \ra \eta \genstar \lambda$ in $n$ steps.
        Then the rule $\hat X \ra \eta$ was added to the CPS grammar
        on an invocation $\mr{CPSProd}(\Pi,\,\tau_\Pi)$
        where $\Pi = X \ra \cdot\,\alpha_1 \ldots \alpha_r$
        and $\eta = \tau_\Pi$.
        By property~$2$, we have $\alpha_1 \ldots \alpha_r \tau_X \genstar \lambda$ in $(n-1)$ steps,
        and hence $X \tau_X \ra \alpha_1 \ldots \alpha_r \tau_X \genstar \lambda$
        in $n$ steps.

        To show property $2$, first consider the case in which the
        dot is at the end. Then $\tau_\Pi = \tau_X$ by definition,
        and since $\alpha_i \ldots \alpha_r$ is the empty string,
        we have $\alpha_i \ldots \alpha_r \tau_X = \tau_X = \tau_\Pi \genstar \lambda$ in $n$~steps, as desired.
        Now consider the case in which the dot is not at the end.
        Let $\Pi' = X \ra \alpha_1 \ldots \alpha_i \cdot \alpha_{i+1} \ldots \alpha_r$.
        We consider the following cases:
        \begin{itemize}
            \item $\alpha_{i}$ is not CPS-triggering.
                Then $\tau_\Pi = \hat\alpha_i \tau_{\Pi'}$.
                Suppose $\lambda = \mu \nu$,
                where $\hat\alpha_i \genstar \mu$ in $n_\mu$ steps,
                and $\tau_{\Pi'} \genstar \nu$ in $n_\nu$ steps,
                with $n = n_\mu + n_\nu$.
                Now clearly $n_\mu, n_\nu \leq n$, and furthermore:
                \begin{itemize}
                    \item If $\alpha_i$ is a terminal symbol, then $|\mu| = 1$, hence $|\nu| < |\lambda|$.
                    \item If $\alpha_i$ is a nonterminal symbol, then $n_\mu > 0$, hence $n_\nu < n$.
                \end{itemize}
                We conclude inductively by property $1$ that
                $\alpha_i = \alpha_i \tau_{\alpha_i} \genstar \mu$
                in $n_\mu$ steps,
                and by property $2$ that
                $\alpha_{i+1} \ldots \alpha_r \tau_X \genstar \nu$ in $n_\nu$ steps,
                and hence $\alpha_i \ldots \alpha_r \tau_X \genstar \mu \nu = \lambda$
                in $n_\mu + n_\nu = n$ steps,
                as desired.
            \item $\alpha_i$ is CPS-triggering, and $\Pi$ is of the form $\alpha_i \ra \gamma \cdot \alpha_i$.
                Then $\tau_\Pi = \hat \alpha_i$,
                and we conclude inductively by property $1$ that
                $\alpha_i \tau_{\alpha_i} \genstar \lambda$ in $n$ steps.
            \item $\alpha_i$ is CPS-triggering, and $\Pi$ is not of the indicated form.
                Then $\tau_{\alpha_i} = \tau_{\Pi'}$,
                and $\tau_\Pi = \hat \alpha_i$.
                We conclude inductively by property $1$ that $\alpha_i \tau_{\Pi'} = \alpha_i \tau_{\alpha_i} \genstar \lambda$
                in $n$ steps.
                Suppose $\lambda = \mu \nu$, where $\alpha_i \genstar \mu$ in $n_\mu$ steps
                and $\tau_{\Pi'} \genstar \nu$ in $n_\nu$ steps,
                with $n = n_\mu + n_\nu$.
                Now clearly $n_\mu, n_\nu \leq n$, and furthermore $n_\mu > 0$ (since $\alpha_i$ is a nonterminal),
                hence $n_\nu < n$.
                Property~$2$ now gives $\alpha_{i+1} \ldots \alpha_r \tau_X \genstar \nu$
                in $n_\nu$ steps,
                and hence $\alpha_i \ldots \alpha_r \tau_X \genstar \mu \nu = \lambda$
                in $n_\mu + n_\nu = n$ steps,
                as desired. \qedhere
        \end{itemize}
\end{proof}

\begin{theorem}[CPS Equivalence]\normalfont
    Let $\GG$ be a flattened grammar,
    and let $\mr{CPS}(\GG)$ denote the CPS-transformed version of $\GG$.
    Then $\GG$ and $\mr{CPS}(\GG)$ generate the same language.
\end{theorem}
\begin{proof}
    Let $S$ denote the start symbol of $\GG$.
    By definition, $S$ is not CPS-eligible and hence not CPS-triggering,
    so $\tau_S = \eps$.
    It follows by Lemma~\ref{lem:cpsfwd} that for every $\lambda \in \Sigma^*$, if $S \genstar \lambda$,
    then $\hat S \genstar \lambda \tau_S = \lambda$.
    Similarly, it follows by Lemma~\ref{lem:cpsrev} that for every $\lambda \in \Sigma^*$,
    if $\hat S \genstar \lambda$, then $S \tau_S = S \genstar \lambda$.
    We conclude that $S$ and $\hat S$ generate identical sets of terminal strings,
    and hence $\GG$ and $\mr{CPS}(\GG)$ generate the same language.
\end{proof}

\noindent We further note that, when a string $X \genstar \lambda$
is generated according to the CPS-transformed grammar,
then when $\hat X \ra \eta \genstar \lambda$
we have $X \tau_X \ra \alpha_1 \ldots \alpha_r \tau_X \genstar \lambda$,
where $X \ra \alpha_1 \ldots \alpha_r$ is the original grammar production
which caused $\hat X \ra \eta$ to be emitted in the $\mr{CPSProd}$ procedure.
We exploit this observation in the LR parsers generated in
our software implementation,
by mapping each CPS production $\hat X \ra \eta$ to its original $X \ra \alpha_1 \ldots \alpha_r$.
Then, during execution, we maintain two stacks:~one for the original AST elements,
and one for the remaining elements from the CPS tail.
When the CPS grammar's
LR automaton calls for a reduction by $\hat X \ra \eta = \eta_1 \ldots \eta_m$,
we look up its original production $X \ra \alpha_1 \ldots \alpha_r$.
If $r \leq m$,
we pop $m$ elements from the first stack,
reduce the earliest $r$ into a result onto the first stack,
and push the remaining unused $(m - r)$ elements onto the second stack;
while if $r > m$,
we pop $r$ elements from the first stack along with $(r - m)$ additional elements from the second stack,
and reduce all $r$ into a result onto the first stack.
This enables us to retain the semantics of the original grammar's AST,
even though the shift/reduce instructions are generated by the CPS grammar's automaton.

\subsection{$k$-follow set partitioning}

\label{sec:optaut}

Given a context-free grammar $\GG$ (e.g., one generated by the CPS procedure of Construction~\ref{cons:cps}),
it remains to construct an LR automaton.
In the case of lookahead $k = 0$,
the standard construction (Construction~\ref{cons:lrknfa}) suffices,
but in the more practical case of $k > 0$,
in order to invoke the construction, we must specify a $k$-follow set partition.
Taking the most refined possible partition
recovers the canonical LR($k$) automaton.
However, this typically leads to an unnecessary proliferation of states,
often making the procedure quite inefficient---a fact which has previously motivated
weaker definitions such as SLR and LALR.
In this work, by contrast, we give a new procedure to compute a $k$-follow set partition
to be used in Construction~\ref{cons:lrknfa},
which retains the full expressive power of canonical LR,
but which typically requires only a small fraction of the number of states.

As discussed in Section~\ref{sec:defs},
we will reason about such $k$-follow set partitions
as equivalence relations over sets of states in the canonical LR NFA
(though we will still use Construction~\ref{cons:lrknfa} to avoid explicitly constructing
the canonical NFA).
Specifically, we will first present algorithms for optimizing arbitrary NFAs by merging states,
then describe how to map such algorithms onto the definition of $k$-follow set partitions.

\begin{definition}[Forward-Admissible Initial Partition]\normalfont
    \label{def:fwdadmiss}
    A partition $L$ of the vertices $V$ of an NFA
    is a forward-admissible initial partition
    if the starting vertex $v_s$ is in a singleton set $\{ v_s \} \in L$.
\end{definition}

\begin{definition}[Forward-Equivalent Vertices]\normalfont
    \label{def:fwdequiv}
    Vertices $v_1,v_2$ are forward-equivalent ($v_1 \sim_{\mr{fwd}} v_2$)
    if for every sequence $\tau = \tau_1 \cdots \tau_k$ with $\tau_i \in A \cup \{\eps\}$,
    $v_1$ is reachable from the starting vertex on $\tau$
    if and only if $v_2$ is reachable from the starting vertex on $\tau$.
\end{definition}

\begin{lemma}[Forward-Equivalent Reachability Preservation]\normalfont
    \label{lem:fwdreachpres}
    Let $\bar N$ be the NFA produced by grouping $N$ by an equivalence relation
    $\sim$, with $\sim$ a refinement of $\sim_{\mr{fwd}}$ (Definition~\ref{def:fwdequiv}).
    Then for every sequence $\tau = \tau_1 \cdots \tau_k$ with $\tau_i \in A \cup \{\eps\}$,
    for every vertex $\bar v$
    reachable on $\tau$ from the starting vertex in $\bar N$,
    for every $v \in \bar v$
    the vertex $v$ is reachable on $\tau$ from the starting vertex in $N$.
\end{lemma}
\begin{proof}
    By induction on $k$.
    Clearly if $k = 0$,
    then $\bar v$ is the starting vertex in $\bar N$,
    and hence there exists $v \in \bar v$ the starting vertex in $N$.
    Since $\bar v$ is an equivalence class under $\sim$,
    and hence a subset of some equivalence class under $\sim_{\mr{fwd}}$,
    it follows that every $v \in \bar v$ is the starting vertex in $N$,
    as desired.
    
    For $k > 0$,
    we have $\tau = \tau_1\cdots \tau_k$ for $\tau_i \in A \cup \{\eps\}$,
    and there exists an edge $\bar v_i \overset{\tau_i}{\ra} \bar v_{i+1}$ for each $i \leq k$.
    The inductive hypothesis gives that every $v_k \in \bar v_k$ is 
    reachable on $\tau_1 \cdots \tau_{k-1}$ from the starting vertex in $N$.
    By definition of $\bar N$, there exists some $v_k \in \bar v_k$, $v_{k+1} \in \bar v_{k+1}$
    such that the edge $v_k \overset{\tau_k}{\ra} v_{k+1}$ exists in $N$;
    hence, this particular $v_{k+1}$ is reachable on $\tau$ from the starting vertex in $N$.
    Since $\bar v_{k+1}$ is an equivalence class under $\sim$,
    and hence a subset of some equivalence class under $\sim_{\mr{fwd}}$,
    it follows that every $v_{k+1} \in \bar v_{k+1}$ is reachable on $\tau$
    from the starting vertex in $N$, as desired.
\end{proof}

\begin{lemma}[Forward-Equivalent Conflict Preservation]\normalfont
    \label{lem:fwdequivconfpres}
    Let $\bar N$ be the NFA produced by grouping $N$ by an equivalence relation
    $\sim$, with $\sim$ a refinement of $\sim_{\mr{fwd}}$ (Definition~\ref{def:fwdequiv}).
    Then there is an NFA conflict in $\bar N$ if and only if there is an NFA conflict in $N$.
\end{lemma}
\begin{proof}
    Clearly if there is a conflict in $N$, then there is a conflict in $\bar N$,
    since $\bar N$ is a coarser grouping than $N$.
    To show the converse, suppose there is a conflict in $\bar N$,
    so that there is a path from $\bar v_0$ to $\bar v_f$ on some sequence $\tau$
    and a path from $\bar v_0$ to $\bar v_f'$ on some sequence $\tau'$
    (where $\tau,\tau'$ yield identical strings when each concatenated),
    such that $\bar v_f$ and $\bar v_f'$ have conflicting actions in $\bar N$.
    Then by definition, there exist $v_f \in \bar v_f$ and $v_f' \in \bar v_f'$
    such that $v_f,v_f'$ have conflicting actions in $N$.
    For such $v_f,v_f'$,
    Lemma~\ref{lem:fwdreachpres}
    shows that $v_f$ and $v_f'$ are each reachable on, resp., $\tau$ and $\tau'$,
    from the starting vertex in $N$.
    Hence there is a conflict in $N$, as desired.
\end{proof}

\begin{definition}[Forward Refinement Algorithm]\normalfont
    \label{def:fwdrefine}
    Fix an initial partitioning $L_0$ of the vertices~$V$.
    Repeat until convergence: for each $Z \in L_i$,
    and each label $\sigma \in A \cup \{\eps\}$,
    define $Z[\sigma]$ to be the set of vertices $z$ for which an edge $z' \overset{\sigma}{\ra} z$ exists
    for some $z' \in Z$.
    Replace $L_i$ with its refinement by each $Z[\sigma]$ in turn,
    obtaining $L_{i+1}$.
    Output the converged partition $L_0, L_1, \ldots \to L$.
\end{definition}

\begin{lemma}[Forward Refinement Correctness]\normalfont
    \label{lem:fwdrefcorr}
    Suppose $L_0$ is a forward-admissible initial partition of the vertices of the NFA~$N$,
    and let $L$ be the result of the forward refinement algorithm~(Definition~\ref{def:fwdrefine})
    on the initial partition $L_0$.
    Fix a sequence $\tau = \tau_1 \cdots \tau_k$
    and let $R(\tau)$ denote the set of vertices reachable on $\tau$ in $N$.
    Then for every set $Z \in L$,
    either $Z \subseteq R(\tau)$ or $Z \cap R(\tau) = \emptyset$.
\end{lemma}
\begin{proof}
    We will show, by induction on $i$,
    that the partition $L_i$ at the $i^{\mr{th}}$ step of the algorithm
    satisfies the corresponding property:
    namely,
    for every sequence $\tau = \tau_1 \cdots \tau_i$,
    for every set $Z \in L_i$,
    either $Z \subseteq R(\tau)$ or $Z \cap R(\tau) = \emptyset$.
    Clearly the claim holds for $i = 0$, since $L_0$ is forward-admissible.
    Suppose $i > 0$, and fix a sequence $\tau = \tau_1 \cdots \tau_i$.
    We may conclude inductively that for every $Z \in L_{i-1}$,
    either $Z \subseteq R(\tau_1 \cdots \tau_{i-1})$ or
    $Z \cap R(\tau_1 \cdots \tau_{i-1}) = \emptyset$.
    In other words, $R(\tau_1 \cdots \tau_{i-1})$ is exactly the union
    of those $Z \in L_{i-1}$ which have nonempty intersection with it.
    On the other hand, the union over such $Z$
    of vertices reachable on label $\tau_i$
    from $Z$ is precisely $R(\tau)$.
    Since $L_i$ has been refined by the image of each such $Z$ under $\tau_i$,
    we conclude that $L_i$ has been refined by sets whose union is precisely $R(\tau)$,
    and the claim follows.
\end{proof}

\begin{lemma}[Forward Refinement Relation]\normalfont
    \label{lem:fwdrefrel}
    Suppose $L_0$ is a forward-admissible initial partition of the vertices of the NFA~$N$,
    and let $L$ be the result of the forward refinement algorithm~(Definition~\ref{def:fwdrefine})
    on the initial partition $L_0$.
    Let $\sim$ be the equivalence relation induced by $L$.
    Then $\sim$ is a refinement of $\sim_{\mr{fwd}}$.
\end{lemma}
\begin{proof}
    Immediate by Lemma~\ref{lem:fwdrefcorr} and the definition of $\sim_{\mr{fwd}}$.
\end{proof}

Lemma~\ref{lem:fwdrefrel} shows that the conditions of Lemma~\ref{lem:fwdequivconfpres}
apply, and hence that for the NFA~$\bar N$ produced by grouping by the partition of the forward
refinement procedure, there will be a conflict if and only if there was a conflict
in the original NFA $N$.
In practice, of course, the grouped NFA $\bar N$ may have significantly fewer states.

To further optimize the NFA,
we also introduce definitions for backward partitioning, analogous to the forward
partitioning definitions given above.

\begin{definition}[Backward-Admissible Initial Partition]\normalfont
    \label{def:bwdadmiss}
    A partition $L$ of the vertices $V$ of the NFA $N$
    is a backward-admissible initial partition
    if, for every lookahead $\mu \in \Gamma^k$,
    and every pair of vertices $v_1,v_2$ with conflicting actions on $\mu$
    reachable on the same sequence $\tau$ from the starting vertex in $N$,
    the following holds.
    Let $\bar v_1,\bar v_2$ be the partition sets
    of, resp., $v_1,v_2$.
    Then for every $v_1' \in \bar v_1$ and every $v_2' \in \bar v_2$,
    the actions on $v_1'$ and on $v_2'$ conflict on lookahead $\mu$.
\end{definition}

We remark that one simple approach to establish a backward-admissible
initial partition, which we will employ in the constructions below,
is to determine the lookahead strings $\mu \in \Gamma^k$ that may potentially be involved in
a conflict at each given dotted production $\hat \Pi$,
and refine the partition of the corresponding NFA vertices
to discriminate appropriately among these lookaheads.
We defer the details to Definition~\ref{def:bwdpartimmlook}, below.

\begin{definition}[Backward-Equivalent Vertices]\normalfont
    \label{def:bwdequiv}
    Fix an initial partition $L_0$ of the vertices of the NFA $N$.
    Vertices $v_1,v_2$ are backward-equivalent with respect to $L_0$ ($v_1 \sim_{\mr{bwd}(L_0)} v_2$)
    if for every $Z \in L_0$,
    and for every sequence $\tau$,
    some element of $Z$ is reachable from $v_1$ on $\tau$
    if and only if some element of $Z$ is reachable from $v_2$ on $\tau$.
\end{definition}

\begin{lemma}[Backward-Equivalent Reachability Preservation]\normalfont
    \label{lem:bwdreachpres}
    Fix an initial partition $L_0$ of the vertices of the NFA $N$.
    Let $\bar N$ be the NFA produced by grouping $N$ by an equivalence relation
    $\sim$, with $\sim$ a refinement of $\sim_{\mr{bwd}(L_0)}$ (Definition~\ref{def:bwdequiv}).
    Then for every $Z \in L_0$, for every vertex $\bar v$ of $\bar N$,
    if $\bar v$ reaches some $\bar w \subseteq Z$ on sequence $\tau = \tau_1 \cdots \tau_k$
    then for every $v \in \bar v$,
    $v$ reaches some element of $Z$ on $\tau$.
\end{lemma}
\begin{proof}
    By induction on $k$.
    Clearly if $k = 0$,
    then $\bar v = \bar w \subseteq Z$,
    so every $v \in \bar v$ reaches $v \in Z$ on $\tau$.
    For $k > 0$,
    we have $\tau = \tau_1\cdots \tau_k$ for $\tau_i \in A \cup \{\eps\}$,
    and there exists an edge $\bar v_i \overset{\tau_i}{\ra} \bar v_{i+1}$ for each $i \leq k$.
    The inductive hypothesis gives that for every $v_2 \in \bar v_2$,
    $v_2$ reaches some element of $Z$ on $\tau_2 \cdots \tau_k$.
    By definition of $\bar N$, there exists some $v_1 \in \bar v_1$, $v_2 \in \bar v_2$
    such that the edge $v_1 \overset{\tau_1}{\ra} v_2$ exists in $N$;
    hence, this particular $v_1$ reaches some element of $Z$ on $\tau$.
    Since $\bar v_1$ is an equivalence class under $\sim$,
    and hence a subset of some equivalence class under $\sim_{\mr{bwd}(L_0)}$,
    it follows that every $v_1 \in \bar v_1$ reaches some element of $Z$ on $\tau$,
    as desired.
\end{proof}

\begin{lemma}[Backward-Equivalent Conflict Preservation]\normalfont
    \label{lem:bwdequivconfpres}
    Let $\bar N$ be the NFA produced by grouping $N$ by an equivalence relation
    $\sim$, with $\sim$ a refinement of $\sim_{\mr{bwd}(L_0)}$ (Definition~\ref{def:bwdequiv}).
    Then there is an NFA conflict in $\bar N$ if and only if there is an NFA conflict in $N$.
\end{lemma}
\begin{proof}
    Clearly if there is a conflict in $N$, then there is a conflict in $\bar N$,
    since $\bar N$ is a coarser grouping than $N$.
    To show the converse, suppose there is a conflict in $\bar N$,
    so that there is a path from $\bar v_0$ to $\bar v_f$ on some sequence $\tau$
    and a path from $\bar v_0$ to $\bar v_f'$ on some sequence $\tau'$
    (where $\tau,\tau'$ yield identical strings when each concatenated),
    such that $\bar v_f \subseteq Z$ and $\bar v_f' \subseteq Z'$
    with $Z,Z' \in L_0$,
    and $\bar v_f, \bar v_f'$ have conflicting actions in $\bar N$.
    Then by Lemma~\ref{lem:bwdreachpres},
    every $v_0 \in \bar v_0$ (including the starting vertex in $N$)
    reaches some $v_f \in Z$ on $\tau$,
    and likewise every $v_0 \in \bar v_0$
    reaches some $v_f' \in Z'$ on $\tau'$.
    By Definition~\ref{def:bwdadmiss},
    the actions on $v_f,v_f'$ conflict,
    i.e., there is a conflict in $N$.
\end{proof}

\begin{definition}[Backward Refinement Algorithm]\normalfont
    \label{def:bwdrefine}
    Fix an initial partition $L_0$ of the vertices~$V$ of the NFA~$N$.
    Repeat until convergence: for each $Z \in L_i$,
    and each label $\sigma \in A \cup \{\eps\}$,
    define $Z[\sigma]$ to be the set of vertices $z$ for which an edge $z \overset{\sigma}{\ra} z'$ exists
    for some $z' \in Z$.
    Replace $L_i$ with its refinement by each $Z[\sigma]$ in turn,
    obtaining $L_{i+1}$.
    Output the converged partition $L_0, L_1, \ldots \to L$.
\end{definition}

\begin{lemma}[Backward Refinement Correctness]\normalfont
    \label{lem:bwdrefcorr}
    Suppose $L_0$ is a partition of the vertices of the NFA~$N$,
    and let $L$ be the result of the backward refinement algorithm~(Definition~\ref{def:bwdrefine})
    on the initial partition $L_0$.
    Fix $Z \in L_0$,
    and let $R_Z(\tau)$ denote the set of vertices that reach some element of $Z$
    on $\tau$ in $N$.
    Then for every set $Y \in L$,
    either $Y \subseteq R_Z(\tau)$ or $Y \cap R_Z(\tau) = \emptyset$.
\end{lemma}
\begin{proof}
    We will show, by induction on $i$,
    that the partition $L_i$ at the $i^{\mr{th}}$ step of the algorithm
    satisfies the corresponding property:
    namely,
    for every sequence $\tau = \tau_1 \cdots \tau_i$,
    for every set $Y \in L_i$,
    either $Y \subseteq R_Z(\tau)$ or $Y \cap R_Z(\tau) = \emptyset$.
    Clearly the claim holds for $i = 0$, since $L_0$ is a partition.
    Suppose $i > 0$, and fix a sequence $\tau = \tau_1 \cdots \tau_i$.
    We may conclude inductively that for every $Y \in L_{i-1}$,
    either $Y \subseteq R_Z(\tau_2 \cdots \tau_i)$ or
    $Y \cap R_Z(\tau_2 \cdots \tau_i) = \emptyset$.
    In other words, $R_Z(\tau_2 \cdots \tau_i)$ is exactly the union
    of those $Y \in L_{i-1}$ which have nonempty intersection with it.
    On the other hand, the union over such $Y$
    of vertices which reach $Y$ on label $\tau_1$
    is precisely $R_Z(\tau)$.
    Since $L_i$ has been refined by the preimage of each such $Y$ under $\tau_1$,
    we conclude that $L_i$ has been refined by sets whose union is precisely $R_Z(\tau)$,
    and the claim follows.
\end{proof}

\begin{lemma}[Backward Refinement Relation]\normalfont
    \label{lem:bwdrefrel}
    Suppose $L_0$ is a partition of the vertices of the NFA~$N$,
    and let $L$ be the result of the backward refinement algorithm~(Definition~\ref{def:bwdrefine})
    on the initial partition $L_0$.
    Let $\sim$ be the equivalence relation induced by $L$.
    Then $\sim$ is a refinement of $\sim_{\mr{bwd}(L_0)}$.
\end{lemma}
\begin{proof}
    Immediate by Lemma~\ref{lem:bwdrefcorr} and the definition of $\sim_{\mr{bwd}(L_0)}$.
\end{proof}

Finally, we show how to adapt the procedures of Definitions~\ref{def:fwdrefine} and \ref{def:bwdrefine}
to operate on the canonical LR($k$) NFA implicitly,
without actually constructing it.
In order to achieve this, we will determine a partition
$L(\hat \Pi)$ of the $k$-follow strings for each dotted production $\hat \Pi$,
and partition the associated set of NFA vertices
$\{(\hat \Pi, \lambda) : \lambda \in \mc{U}(\hat \Pi)\}$
according to $L(\hat \Pi)$ independently for each~$\hat \Pi$.
This means that our NFA vertex partitions may be more refined than strictly necessary,
since they will by construction separate the vertices of distinct dotted productions $\hat \Pi,\hat \Pi'$.
However, by making this simplification, we will avoid constructing the full canonical LR($k$) NFA,
resulting in a significantly more efficient NFA optimization procedure.

To begin with, we define the set of all strings $\lambda \in \Gamma^k$
which can possibly follow each production~$\Pi$.
Our definition follows the standard SLR flood-fill algorithm.
\begin{definition}[Forward-Reachable $k$-Follow Strings]\normalfont
    \label{def:fwdreachlook}
    For each production $\Pi$ of the grammar $\GG$,
    we define the set of forward-reachable $k$-follow strings $\lambda \in \Gamma^k$
    of $\Pi$ iteratively as follows.
    \begin{itemize}
        \item The string $\srt^k$ is forward-reachable at
            the starting production $\Pi = S \ra \eta$.
        \item For productions $\Pi = X \ra \alpha Y \beta$
            and $\Pi' = Y \ra \gamma$,
            if $\lambda$ is forward-reachable at $\Pi$,
            then for every $\mu \in \fst_k(\fstgen(\beta \lambda))$,
            $\mu$ is forward-reachable at $\Pi'$.
    \end{itemize}
\end{definition}

For ease of exposition, we also introduce a notion of lookaheads {\em compatible}
with a given $k$-follow string (for a given dotted production):
\begin{definition}[Compatible Lookaheads]\normalfont
    \label{def:lacompat}
    Let $\hat \Pi = X \ra \alpha \cdot \beta$ be a dotted production.
    A lookahead $\mu \in \Gamma^k$ is {\em $\hat \Pi$-compatible} with
    a $k$-follow string $\lambda \in \Gamma^k$
    if $\mu \in \fst_k(\fstgen(\beta \lambda))$.
\end{definition}

We now define the set of lookaheads $\mu \in \Gamma^k$
that are potentially involved in a conflict, for each dotted production $\hat \Pi$.
\begin{definition}[Potentially-Conflicting Lookaheads]\normalfont
    \label{def:potconflook}
    For a dotted production $\hat \Pi$,
    a lookahead $\mu \in \Gamma^k$ is {\em potentially-conflicting}
    if there exists a dotted production $\hat \Pi'$
    and $k$-follow strings $\lambda \in \mc{U}(\hat \Pi)$, $\lambda' \in \mc{U}(\hat \Pi')$
    such that the following conditions hold:
    \begin{itemize}
        \item The vertices $(\hat \Pi, \{\eps\})$ and $(\hat \Pi', \{\eps\})$
            are reachable on the same string $\zeta \in (\Sigma \cup \Lambda)^*$
            in the LR($0$)~NFA.
        \item The vertices $(\hat \Pi, \{\lambda\})$ and $(\hat \Pi', \{\lambda'\})$
            have conflicting actions on lookahead $\mu$ in the canonical LR($k$)~NFA.
    \end{itemize}
\end{definition}

\noindent Given these potentially-conflicting lookaheads, we can now form a partition
of the $k$-follow sets of each dotted production that enforces the appropriate separations.
\begin{definition}[Immediately-Conflicting Backward Partition]\normalfont
    \label{def:bwdpartimmlook}
    The immediately-conflicting backward partition $L_0$ is defined as follows.
    For each dotted production $\hat \Pi$,
    let the partition $L_0(\hat \Pi)$ be defined by an equivalence relation $\equiv_{\hat \Pi}$
    as follows.
    For $k$-follow strings $\lambda,\lambda' \in \Gamma^k$,
    we have $\lambda \equiv_{\hat \Pi} \lambda'$ if, for every potentially-conflicting lookahead $\mu$
    for $\hat \Pi$,
    the lookahead $\mu$ is $\hat \Pi$-compatible with $\lambda$
    if and only if $\mu$ is $\hat \Pi$-compatible with $\lambda'$ (Definition~\ref{def:lacompat}).
\end{definition}

\begin{lemma}\normalfont
    \label{lem:bwdpartimmlookadmiss}
    The partition $L_0$ given by Definition~\ref{def:bwdpartimmlook} is a backward-admissible
    initial partition~(Definition~\ref{def:bwdadmiss}).
\end{lemma}
\begin{proof}
    Fix a pair of vertices $v_1 = (\hat \Pi_1, \lambda_1), v_2 = (\hat \Pi_2, \lambda_2)$
    with conflicting actions on some lookahead $\mu \in \Gamma^k$,
    and fix $\lambda_1' \equiv_{\hat \Pi_1} \lambda_1$ and $\lambda_2' \equiv_{\hat \Pi_2} \lambda_2$.
    By definition of the canonical LR($k$)~NFA,
    we have that $\lambda_1$ is $\hat \Pi_1$-compatible with $\mu$,
    and that $\lambda_2$ is $\hat \Pi_2$-compatible with $\mu$.
    Moreover, since $v_1,v_2$ are reachable on the same string $\zeta$ in the canonical LR($k$) NFA,
    we have $(\hat \Pi_1, \{\eps\}),(\hat \Pi_2, \{\eps\})$ each reachable on $\zeta$ in the LR($0$) NFA,
    and thus $\mu$ is a potentially-conflicting lookahead (Definition~\ref{def:potconflook}).
    By Definition~\ref{def:bwdpartimmlook} we conclude that $\lambda_1'$ is $\hat \Pi_1$-compatible with $\mu$,
    and $\lambda_2'$ is $\hat \Pi_2$-compatible with $\mu$.
    Hence the actions on $(\hat \Pi_1, \lambda_1'),(\hat \Pi_2, \lambda_2')$ conflict on $\mu$, as desired.
\end{proof}

Since Definition~\ref{def:bwdpartimmlook} gives a backward-admissible partition,
the procedure of Definition~\ref{def:bwdrefine}
applies; moreover, such a procedure can be executed without constructing the full
LR($k$) automaton, by maintaining for each dotted production $\hat \Pi$ the current partitioning of
the $k$-follow set on $\hat \Pi$ at each point during the procedure.
Evidently, the forward procedure (Definition~\ref{def:fwdrefine})
can be executed in a similar manner, resulting in the following algorithm.

\begin{construction}[Forward/Backward $k$-Follow Set Partition Algorithm]\normalfont
    \label{cons:fwdbwdalg}
    Let $\GG$ be a context-free grammar, let $N_0$ denote its LR($0$) NFA,
    and let $N$ denote its canonical LR($k$)~NFA.
    Perform the following steps.
    \begin{enumerate}
        \item Compute the forward-reachable $k$-follow set (Definition~\ref{def:fwdreachlook})
            for each dotted production~$\hat \Pi$.
        \item Form the LR($0$)~DFA by invoking the standard subset construction on $N_0$.
            Using the DFA to determine reachability on strings $\zeta \in (\Lambda \cup \Sigma)^*$,
            determine the potentially-conflicting lookaheads (Definition~\ref{def:potconflook})
            for each dotted production $\hat \Pi$.
        \item Construct the immediately-conflicting backward partition $L_{0,\mr{bwd}}$
            of the vertices in $N$ (Definition~\ref{def:bwdpartimmlook}),
            maintaining the partition implicitly via partitioning the set of forward-reachable
            $k$-follow strings for each production.
        \item Execute the backward refinement algorithm (Definition~\ref{def:bwdrefine})
            on the partition $L_{0,\mr{bwd}}$
            (again operating implicitly over the partitions of $k$-follow strings,
            rather than constructing the canonical NFA explicitly).
            We write $L_{\mr{bwd}}(\hat \Pi)$ to denote
            the resulting partition of $k$-follow strings for each production $\hat \Pi$.
        \item Taking the constituent subsets $Z \in L_{\mr{bwd}}(\hat \Pi)$
            as the new base elements,
            form a trivial partition $L_{0,\mr{fwd}}(\hat \Pi)$ over these elements,
            consisting of a single subset containing all subsets $Z \in L_{\mr{bwd}}(\Pi)$
            for each dotted production $\hat \Pi$.
            Since the new partition $L_{0,\mr{fwd}}(\hat \Pi)$ still already separates vertices of
            distinct dotted productions $\hat \Pi \neq \hat \Pi'$,
            it is forward-admissible (Definition~\ref{def:fwdadmiss}).
        \item Execute the forward refinement algorithm (Definition~\ref{def:fwdrefine})
            on the partition $L_{0,\mr{fwd}}$
            (again operating implicitly over the partitions of sets of $k$-follow strings,
            rather than constructing the canonical NFA).  
            We write $L(\hat \Pi)$ to denote the resulting partition of $k$-follow strings
            for each production~$\hat \Pi$.
        \item Output the LR($k$)~NFA produced by invoking Construction~\ref{cons:lrknfa}
            on the final partition $L(\hat \Pi)$.
    \end{enumerate}
\end{construction}

\begin{theorem}[Forward/Backward Partition Correctness]\normalfont
    Let $\GG$ be a context-free grammar, let $N$ denote its canonical LR($k$)~NFA,
    and let $N^*$ denote the NFA formed by running the forward/backward algorithm~(Construction~\ref{cons:fwdbwdalg})
    on $\GG$.
    Then there is a conflict in $N^*$ if and only if there is a conflict in $N$.
\end{theorem}
\begin{proof}
    Immediate from Lemmas~\ref{lem:fwdequivconfpres} and \ref{lem:bwdequivconfpres}.
\end{proof}

We note that other combinations of forward- and backward-partitioning are possible;
however, empirically we find that a single backward pass followed by a single forward pass
(i.e., the procedure of Construction~\ref{cons:fwdbwdalg})
suffices to produce remarkably small automata in practice---the backward pass easily zeroes in on
the ``problematic'' $k$-follow strings which must be distinguished from their complements,
and the forward pass groups $k$-follow strings which are ``similar'' (e.g.,
binary operators with identical precedence).

\subsection{Conflict tracing}

Debuggability is crucial for practical automatic parser generation.
In existing parser generators, the source of a conflict (shift/reduce, reduce/reduce)
is often very difficult to diagnose,
and new conflicts can easily arise from seemingly incidental changes to the BNF grammar.
To remedy this, we introduce a method of tracing conflicts in the LR automata,
which enables the reconstruction of a minimal ``confusing input'' which triggers the conflict.

To begin with, it will be useful to precompute, for each symbol of the grammar,
a shortest string generated by that symbol.
This can be done via a standard dynamic programming approach.
\begin{construction}[Generated String Precomputation]\normalfont
    \label{cons:genstrpre}
    For a grammar $\GG$ over nonterminals $\Lambda$ and terminals $\Sigma$,
    initialize a table as follows.
    For terminals $\sigma \in \Sigma$,
    initialize $\sigma \mapsto (1, \sigma)$,
    and for nonterminals $\alpha \in \Lambda$, initialize $\alpha \mapsto (\infty, \bot)$.
    For each production $\Pi = X \ra \cdots \sigma \cdots$
    in which each terminal $\sigma$ appears in the right-hand side
    (and $\Pi$ is not already in the queue),
    enqueue $\Pi$.
    Then, while the queue is not empty, repeat:
    \begin{enumerate}
        \item Dequeue a production $\Pi = X \ra \eta_1 \ldots \eta_r$.
        \item Suppose we currently have $\eta_i \mapsto (\ell_i, \lambda_i)$ in the table, for each $i \in [r]$.
            Compute $\ell = \sum \ell_i$, and $\lambda = \lambda_1 \cdots \lambda_r$
            (where $\bot$ concatenated with any string is $\bot$).
        \item If $\ell$ is strictly less than the current length entry for $X$ in the table,
            then replace the current entry with $X \mapsto (\ell, \lambda)$,
            and for each production $\Pi'$ in which $X$ appears in the right-hand side
            (and $\Pi'$ is not already in the queue),
            enqueue $\Pi'$.
    \end{enumerate}
\end{construction}

\noindent Similarly, we can compute a shortest string generated by a given string
(of terminals and/or nonterminals)
which begins with a given prefix.
We begin by tabulating these values for each symbol, then proceed to the case of general strings.
\begin{definition}[Prefix Table]\normalfont
    \label{def:preftable}
    Fix a grammar $\GG$ over nonterminals $\Lambda$ and terminals $\Sigma$
    and an integer $k \geq 0$.
    A $k$-prefix table for a string $\zeta \in (\Lambda \cup \Sigma)^*$
    consists of a mapping $\Sigma^{\leq k} \to \Sigma^*$
    such that for a given $\alpha \in \Sigma^{\leq k}$,
    the key $\alpha$ is present if and only if there exists $\beta \in \Sigma^*$
    such that $\zeta \genstar \beta$ and $\alpha$ is the truncation to a prefix of
    at most $k$ symbols of $\beta$;
    and, for such $\alpha$ that are present, we have $\alpha \mapsto \beta^*$,
    where $\beta^*$ is a shortest such $\beta$.
\end{definition}

\begin{construction}[Prefix Table Precomputation]\normalfont
    \label{cons:preftableprecomp}
    Fix a grammar $\GG$ over nonterminals $\Lambda$ and terminals $\Sigma$
    and an integer $k > 0$.
    The output of the construction will be a $k$-prefix table (Definition~\ref{def:preftable})
    for each symbol of $(\Lambda \cup \Sigma)$,
    as well as a $k$-prefix table for the tail of each dotted production of $\GG$
    (i.e., for the string $\beta$ for each dotted production $X \ra \alpha \cdot \beta$).
    The construction proceeds as follows.
    \begin{enumerate}
        \item For each terminal $\sigma \in \Sigma$, initialize the $k$-prefix table
            to $\{ \sigma \mapsto \sigma \}$, and enqueue the symbol $\sigma$.
        \item For each nonterminal $\zeta \in \Lambda$, initialize the $k$-prefix table to $\{ \}$.
        \item For each dotted production tail $X \ra \alpha \cdot \beta$
            such that $\beta \neq \eps$,
            initialize the $k$-prefix table to $\{ \}$.
        \item For each dotted production tail $X \ra \beta\,\cdot$,
            initialize the $k$-prefix table to $\{ \eps \mapsto \eps \}$,
            and enqueue the dotted production tail $X \ra \beta\,\cdot$.
        \item While the queue is not empty:
        \begin{itemize}
            \item Dequeue an item (a symbol or dotted production tail).
                If the item is a symbol $\zeta$, then enqueue every dotted production tail
                of the form $X \ra \alpha \zeta \cdot \beta$ and continue from the top of the loop.
                If the item is a dotted production tail with the dot at the beginning,
                i.e., $\hat \Pi = X \ra \cdot\,\alpha$,
                then for each entry $\gamma \mapsto \delta$ in the $k$-prefix table for $\hat \Pi$,
                add $\gamma \mapsto \delta$ to the $k$-prefix table for $X$
                (unless $\gamma$ is already present and maps to a string at least as short as~$\delta$);
                enqueue $X$ if this process resulted in any new entries in its $k$-prefix table; and
                continue from the top of the loop.
                Henceforth we assume the item is a dotted production tail
                such that the dot is not at the beginning, i.e., $\hat \Pi = X \ra \alpha \zeta \cdot \beta$
                for some symbol $\zeta$.
            \item For each entry $\gamma \mapsto \delta$ in the $k$-prefix table for $\zeta$,
                and each entry $\gamma' \mapsto \delta'$ in the table for~$\hat \Pi$,
                form the concatenation mapping $\hat \gamma \mapsto \hat \delta$,
                where $\hat \gamma$ is the truncation of $\gamma \gamma'$ to a prefix of at most $k$,
                and $\hat \delta = \delta \delta'$.
                For each such entry $\hat \gamma \mapsto \hat \delta$,
                add it to the $k$-prefix table for $\hat \Pi' = X \ra \alpha \cdot \zeta \beta$
                (unless $\hat \gamma$ is already present and maps to a string at least as short
                as $\hat \delta$).
                If this process resulted in any new entries in the $k$-prefix table for $\hat \Pi'$,
                then enqueue~$\hat \Pi'$.
        \end{itemize}
    \end{enumerate}
\end{construction}

\begin{construction}[Prefix-Matching Generated String Algorithm]\normalfont
    \label{cons:prefmatch}
    Fix a grammar $\GG$ over nonterminals $\Lambda$ and terminals $\Sigma$
    (where $\Gamma = \Sigma \cup \{\srt\}$),
    an integer $k \geq 0$ and a target lookahead $\lambda = \lambda_1 \ldots \lambda_k \in \Gamma^k$,
    and an input string $\zeta = \zeta_1 \ldots \zeta_s \in (\Lambda \cup \Gamma)^*$.
    The output of the construction will be a shortest string generated by $s$
    that begins with the prefix $\lambda$, if one exists.

    To compute such a string, perform a shortest-path search in the following graph.
    The vertices of the graph are pairs $(i, j)$ for $i \in [0,k], j \in [0,s]$.
    The initial vertex is $(0, 0)$, and the target vertex is $(k, s)$.
    There is a edge from $(i, j)$ to $(i', j+1)$ of length $\ell$ if there
    is an entry $\alpha \mapsto \beta$ in the $k$-prefix table for
    $\zeta_{j+1}$ (Definition~\ref{def:preftable})
    such that either $\alpha = \beta = \lambda_{i+1} \ldots \lambda_{i'}$,
    or $i' = k$ and $\alpha$ begins with the prefix $\lambda_{i+1} \ldots \lambda_{i'}$;
    and $\beta$ has length $\ell$.
\end{construction}

Equipped with the above constructions, we now present the full conflict tracing algorithm.

\begin{construction}[LR Conflict Tracing Algorithm]\normalfont
    \label{cons:conftrace}
    Let $N$ be an LR($k$)~NFA for the grammar~$\GG$,
    and let $D$ be its corresponding DFA.
    For each vertex $\bar v$ of $D$ with conflicting accept actions $a,a' \in A$,
    perform the following steps to trace the conflict.
    \begin{enumerate}
        \item Perform a shortest-path search in $D$, from the starting vertex
            to the conflicting vertex $\bar v$,
            where the length of an edge with label $\alpha \in \Lambda \cup \Sigma$
            is defined to be the length of the shortest string generated by $\alpha$
            (Construction~\ref{cons:genstrpre}).
            Let $\zeta_1 \cdots \zeta_r$ denote the labels on the edges of the resulting path.
        \item Perform a shortest-path search in $N \times \{0,\ldots,r\}$,
            where the starting vertex is $(v, 0)$ for $v$ the starting vertex of $N$,
            and the edges are defined as follows.
            \begin{itemize}
                \item If there is an edge $v \overset{\zeta_i}{\to} w$ in $N$,
                    then there is an edge $(v,\,i-1) \to (w,\,i)$ in the product graph
                    with length $0$.
                \item If there is an edge $v \overset{\eps}{\to} w$ in $N$,
                    then for all $i$ there is an edge $(v,\,i) \to (w,\,i)$ in the product graph,
                    each with length $\ell$, where $\ell$ is the sum of lengths of the shortest strings
                    generated by the tail of the production on $w$
                    (i.e., if $w = (X \ra \alpha \cdot \beta,\,\bar\lambda)$,
                    then $\ell$ is the sum of the lengths generated by the symbols in $\beta$).
            \end{itemize}
            Let $v_0,\ldots,v_s$ denote a resulting NFA path from the starting vertex
            to a vertex $(v,\,r)$ whose accept action is~$a$ on lookahead $\lambda$,
            and $v'_0,\ldots,v'_{s'}$ denote a path to a vertex $(v',\,r)$
            whose accept action is $a'$ on lookahead $\lambda$.
        \item Construct the string $\omega = \zeta_1 \ldots \zeta_r \beta_s \beta_{s-1} \ldots \beta_0$,
            where $\beta_j$ is defined as follows:
            \begin{itemize}
                \item If $j < s$ and the edge $v_j \to v_{j+1}$ is an $\eps$-edge,
                    then $\beta_j$ is the proper tail of the production of $v_j$
                    (i.e., $\beta$ if the production of $v_j$ is some $\Pi = \alpha \cdot Y \beta$).
                \item If $j < s$ and the edge $v_j \to v_{j+1}$ is not an $\eps$-edge,
                    then $\beta_j = \eps$.
                \item If $j = s$, then $\beta_j$ is the tail of the production of $v_j$
                    (i.e., $\beta$ if the production of $v_j$ is some $\Pi = \alpha \cdot \beta$).
            \end{itemize}
            Also construct the string $\omega' = \zeta_1 \ldots \zeta_r \beta'_s \beta'_{s-1} \ldots \beta'_0$,
            where $\beta'_j$ is defined analogously to $\beta_j$, but for $v'_i$.
        \item Construct the string $\hat \omega$ (resp., $\hat \omega'$) from $\omega$
            (resp., $\omega'$)
            as follows.
            Substitute each nonterminal in $\zeta_1 \ldots \zeta_r$ with a minimal string
            that it generates~(Construction~\ref{cons:genstrpre}), producing $\hat \zeta$.
            Then, using Construction~\ref{cons:prefmatch},
            find a string $\hat \beta$ (resp., $\hat \beta'$)
            generated by $\beta_s \ldots \beta_0 \srt^k$
            (resp., $\beta_s' \ldots \beta_0' \srt^k$)
            which begins with the given conflicting lookahead $\lambda$
            (its existence follows inductively from the definition of the lookahead sets in
            Construction~\ref{cons:lrknfa}).
            Define $\hat \omega = \hat \zeta \hat \beta$ (resp., $\hat \omega' = \hat \zeta \hat \beta'$).
        \item Output the pair of ``confusing inputs'' $\hat \omega,\,\hat \omega'$.
    \end{enumerate}
\end{construction}

The ``confusing inputs'' $\hat \omega,\, \hat \omega'$ output
by Construction~\ref{cons:conftrace} are confusing in the following sense.
First, evidently they are both generated by the grammar,
and have a common prefix $\hat \zeta$.
Yet, at the point when the LR parser has processed the prefix $\hat \zeta$,
there exist two different possible completions ($\hat \omega, \hat \omega'$),
both with lookahead $\lambda \in \Gamma^k$,
such that the LR automaton calls for distinct shift/reduce actions
at that point in the string.
In this sense, a ``confusing input pair'' generalizes the concept of an ambiguity in the grammar.
If we had $\hat \omega = \hat \omega'$ (which can indeed arise in ambiguous grammars),
then the fact that $\hat \zeta$ calls for conflicting actions would imply two distinct
parse trees for the string $\hat \omega = \hat \omega'$.
In the more general case, however, a confusing input need not indicate an ambiguity;
only an LR($k$) conflict.
Empirically, we have found that in translating real industrial grammars,
this conflict tracing method almost always pinpoints the precise cause of LR conflicts,
and the ``confusing inputs'' are indeed confusing to a human observer.

One minor technical issue remains when we employ Construction~\ref{cons:conftrace} in practice.
While the existence of a completion $\hat \beta$ matching the given lookahead
$\beta_s \ldots \beta_0 \srt^k$
is guaranteed in the canonical LR($k$) automaton,
it is not obvious that this also holds in the LR($k$) automaton resulting from
the forward/backward algorithm~(Construction~\ref{cons:fwdbwdalg}).
To show this, we can generalize the results of the previous section
to a modified LR NFA in which $\eps$-edges are replaced by special symbols
$\eps_{\hat \Pi, \hat \Pi'}$ that indicate which specific dotted production is predicted
when traversing the edge.
The more general analogues of Lemmas~\ref{lem:fwdreachpres} and \ref{lem:bwdreachpres}
then establish the existence of paths in the canonical LR($k$) NFA,
which match not only the symbol sequence but also the precise sequence of predicted productions,
and hence the desired tail $\beta_s \ldots \beta_0 \srt^k$.

\subsection{Nonterminal attributes}
\label{sec:attr}

In parsing industrial programming languages, we often find a need for
non-context-free properties of nonterminals:~either semantic attributes,
such as ``expression in a type context''
versus ``expression in a value context'',
or syntactic attributes, such as are needed in Golang's greedy parsing of expressions
like ``\texttt{if * func () \{\}}''.
To address these requirements, we give a construction which permits nonterminals
to be accompanied by {\em attributes}, which
can be either binary-valued (as in most syntactic attributes),
or integer-valued (as in the case of operator precedence).
Relationships between attributes are then represented by constraints on grammar rules.

We begin by defining the possible domains for nonterminal attributes,
and proceed to define the syntax of attribute constraints on BNF grammar rules.
\begin{definition}[Attribute Value Type]\normalfont
    \label{def:attrvaltype}
    An attribute value type is either boolean (i.e., $\{0, 1\}$)
    or integer-valued with range $n$ (i.e., $\{0,1,\ldots,n-1\}$).
\end{definition}
\begin{definition}[Nonterminal Attribute Domain]\normalfont
    Each nonterminal symbol has an {\em attribute domain}
    consisting of a mapping from attribute keys (fresh symbols $\alpha$ in some set $A$)
    to attribute value types (Definition~\ref{def:attrvaltype}).
\end{definition}
\begin{definition}[BNF Attribute Constraints]\normalfont
    \label{def:bnfattrconstr}
    Each rule of a BNF grammar is accompanied by a set of {\em attribute constraints},
    each of one of the following forms.
    \begin{itemize}
        \item $\mr{lhs}[\alpha] \leq b$
        \item $\mc{R}[\alpha] \geq b$
        \item $\mr{lhs}[\alpha] \leq \mc{R}[\beta]$
    \end{itemize}
    where $\alpha,\beta \in A$ are attribute keys,
    $b$ is an value in the attribute domain of the corresponding symbol,
    and $\mc{R}$ takes one of the following values:
    \begin{itemize}
        \item ``$\mr{rhs}$'', referring to every subexpression of the right-hand side of the rule.
        \item ``$\mr{rhs\_begin}$'', referring to subexpressions which are (syntactically)
            the first in the right-hand side of the rule.
        \item ``$\mr{rhs\_end}$'', referring to subexpressions which are (syntactically)
            the last in the right-hand side of the rule.
        \item ``$\mr{rhs\_mid}$'', referring to subexpressions which are (syntactically)
            neither first nor last in the right-hand side of the rule.
        \item ``$\mr{rhs\_tag}_{[\tau]}$'', referring to subexpressions which are explicitly tagged
            in the BNF syntax with the symbol $\tau$.
    \end{itemize}
\end{definition}
\begin{definition}[Flattened Attribute Constraints]\normalfont
    \label{def:flatattrconstr}
    Each production $\Pi = X \ra \alpha_1 \ldots \alpha_r$
    of a flattened (context-free) grammar is accompanied by a set of {\em flattened attribute constraints},
    each of one of the following forms.
    \begin{itemize}
        \item $\mr{lhs}[\alpha] \leq b$
        \item $\mr{rhs}_i[\alpha] \geq b$
        \item $\mr{lhs}[\alpha] \leq \mr{rhs}_i[\beta]$
    \end{itemize}
    where $i \in [r]$ refers to an index into the right-hand side of the production,
    and as above, $\alpha,\beta$ are attribute keys and $b$ is an attribute
    in the domain of the corresponding symbol.
\end{definition}

It is straightforward to adapt the grammar flattening procedure (Construction~\ref{cons:grmflat})
and LR automaton definition (Construction~\ref{cons:lrknfa}) to incorporate attribute constraints
of the form described in Definitions~\ref{def:bnfattrconstr} and \ref{def:flatattrconstr};
we only mention a few technical details.
First, when a grammar in BNF form is flattened to a context-free grammar,
fresh symbols generated during the translation of the right-hand side inherit
the attribute domains of their left-hand side (and propagate attributes via constraints
accordingly)---e.g., in the flattening of a rule $E := (A\,|\,B)\,C$,
a fresh symbol $X$ is generated with $E \ra X\,C$, $X \ra A$, and $X \ra B$;
and as a result, $X$ inherits the attribute domain of $E$, and the production $E \ra X\,C$
carries the constraint $\mr{lhs}[\alpha] \leq \mr{rhs}_1[\alpha]$
for every attribute key $\alpha$ in that domain.
This enables attributes to be carried through, e.g., via ``$\mr{rhs\_begin}$'' to $A$ or to $B$.
Note, however, that attributes are not carried through to right-hand side items
when they are not in those items' domain; thus, for instance, a rule $E := x\,A$
where $x$ is a terminal symbol implies no constraint on $x$,
even if the BNF rule carries a $\mr{rhs\_begin}$ constraint,
since the attribute domain of a terminal symbol is empty by definition.

Second, when we form the LR($k$) automata in the presence of attribute constraints,
NFA vertices may carry a set of attribute bounds of the form $\mr{lhs}[\alpha] \geq b$,
with each distinct set of attribute bounds forming a separate vertex.
Because attribute constraints are separable and monotonic,
a set of bounds of this form suffices to determine, in turn, the bounds
on each of the symbols of the right-hand side,
which themselves then become $\mr{lhs}$ constraints in the vertices predicted.
Moreover, the sets of bounds that arise in practice remain quite manageable,
owing to the fact that most constraints do not influence the predicted right-hand side
and can hence be omitted---e.g., for a vertex $(X \ra A \cdot B, -)$,
a bound $\mr{lhs}[\alpha] \geq b$ has no influence, and can be dropped from the vertex,
unless there is a constraint of the specific form $\mr{lhs}[\alpha] \ra \mr{rhs}_2[\beta]$.
(Even a constraint of the form $\mr{lhs}[\alpha] \ra \mr{rhs}_1[\beta]$ has no effect,
since the dot has already advanced past the first symbol of the right-hand side.)
Once the LR($k$) NFA has been formed with these attribute bounds,
the DFA can then be formed via a modified subset construction:~as usual,
we use subsets of the NFA vertices,
but the edges between them consist of concrete attribute values
(rather than attribute bounds)---e.g., there may be an edge whose label
is some symbol $\zeta[\alpha=0,\,\beta=2,\,\gamma=1]$ where $\alpha,\beta,\gamma$ are attribute keys,
and the values are consistent with the constraints on the production in question.
This approach has the added benefit that only {\em inhabited} symbols $\zeta[\alpha,\beta,\gamma]$,
i.e., only those attribute values which are actually attainable by the grammar,
need be added as labels in the DFA.

In conjunction with the techniques described in other sections,
this approach to expressing attribute constraints suffices to describe most
semantic and syntactic properties of interest in real programming language grammars.
For boolean-valued attributes (e.g., ``type expression'', ``value expression''),
the translation under this paradigm is clear:~one may introduce attributes,
e.g., $\mr{T}$ and $\mr{V}$ for type expressions and value expressions, respectively;
and write, e.g., ``ArrayTypeExpr := $\mr{Expr}_{[\tau]}\,[\,\mr{Expr}_{[\omega]}\,]$'',
with constraints such as ``$\lnot \mr{lhs}[V]$'', ``$\mr{rhs\_tag}_{[\tau]}[T]$'',
and ``$\mr{rhs\_tag}_{[\omega]}[V]$''.
For attributes such as operator precedence,
the translation is somewhat more involved and requires
general integer-valued attributes.
We describe one such translation here.

\begin{definition}[BNF Precedence Stanza]\normalfont
    \label{def:bnfprecstanza}
    A precedence stanza for a BNF grammar is a sequence of items
    of the form $(\{\Pi_1,\ldots,\Pi_s\}, \mc{P})$,
    where $\{\Pi_1,\ldots,\Pi_s\}$ is a set of BNF grammar rules,
    and $\mc{P}$, an associativity specification,
    is one of $\{\mr{none},\, \mr{left},\, \mr{right},\, \mr{prefix},\, \mr{postfix}\}$.
    We require each BNF rule $\Pi$ to appear in at most one set $\{\Pi_1,\ldots,\Pi_s\}$,
    and we define the precedence level of $\Pi$ to be the zero-based index of its containing set within
    the full sequence of the precedence stanza (if it appears).
    We also require that for every nonterminal symbol $E$ in the BNF,
    if any one of the productions $E := \ldots$ has a precedence level,
    then all productions $E := \ldots$ have a precedence level.
\end{definition}
\begin{construction}[BNF Precedence Constraints]\normalfont
    Given a BNF precedence stanza (Definition~\ref{def:bnfprecstanza}),
    we define associated attributes and constraints as follows.
    First, we define two integer-valued attributes $\mr{prL},\mr{prR}$
    (intuitively, the left- and right-precedence),
    whose range is $\{0,\ldots,n-1\}$,
    where $n$ is the number of precedence levels (i.e., number of items in the precedence stanza
    sequence).
    Then, for each BNF rule $E := \ldots$ which appears in the precedence stanza
    at precedence level $\ell$, we add the following constraints
    depending on its associativity specification:
    \begin{itemize}
        \item $\mr{none}$:
            \begin{itemize}
                \item $\mr{lhs}[\mr{prL}] \leq \ell$
                \item $\mr{lhs}[\mr{prR}] \leq \ell$
                \item $\mr{rhs}[\mr{prL}] \geq \ell + 1$
                \item $\mr{rhs}[\mr{prR}] \geq \ell + 1$
            \end{itemize}
        \item $\mr{left}$:
            \begin{itemize}
                \item $\mr{lhs}[\mr{prL}] \leq \ell$
                \item $\mr{lhs}[\mr{prR}] \leq \ell$
                \item $\mr{rhs\_begin}[\mr{prR}] \geq \ell$
                \item $\mr{rhs\_mid}[\mr{prL}] \geq \ell + 1$
                \item $\mr{rhs\_mid}[\mr{prR}] \geq \ell + 1$
                \item $\mr{rhs\_end}[\mr{prL}] \geq \ell + 1$
                \item $\mr{lhs}[\mr{prL}] \leq \mr{rhs\_begin}[\mr{prL}]$
                \item $\mr{lhs}[\mr{prR}] \leq \mr{rhs\_end}[\mr{prR}]$
            \end{itemize}
        \item $\mr{right}$:
            \begin{itemize}
                \item $\mr{lhs}[\mr{prL}] \leq \ell$
                \item $\mr{lhs}[\mr{prR}] \leq \ell$
                \item $\mr{rhs\_begin}[\mr{prR}] \geq \ell + 1$
                \item $\mr{rhs\_mid}[\mr{prL}] \geq \ell + 1$
                \item $\mr{rhs\_mid}[\mr{prR}] \geq \ell + 1$
                \item $\mr{rhs\_end}[\mr{prL}] \geq \ell$
                \item $\mr{lhs}[\mr{prL}] \leq \mr{rhs\_begin}[\mr{prL}]$
                \item $\mr{lhs}[\mr{prR}] \leq \mr{rhs\_end}[\mr{prR}]$
            \end{itemize}
        \item $\mr{prefix}$:
            \begin{itemize}
                \item $\mr{lhs}[\mr{prR}] \leq \ell$
                \item $\mr{rhs\_mid}[\mr{prL}] \geq \ell + 1$
                \item $\mr{rhs\_mid}[\mr{prR}] \geq \ell + 1$
                \item $\mr{rhs\_end}[\mr{prL}] \geq \ell + 1$
                \item $\mr{lhs}[\mr{prR}] \leq \mr{rhs}[\mr{prR}]$
            \end{itemize}
        \item $\mr{postfix}$:
            \begin{itemize}
                \item $\mr{lhs}[\mr{prL}] \leq \ell$
                \item $\mr{rhs\_begin}[\mr{prR}] \geq \ell + 1$
                \item $\mr{rhs\_mid}[\mr{prL}] \geq \ell + 1$
                \item $\mr{rhs\_mid}[\mr{prR}] \geq \ell + 1$
                \item $\mr{lhs}[\mr{prL}] \leq \mr{rhs}[\mr{prL}]$
            \end{itemize}
    \end{itemize}
\end{construction}
We note that both left- and right-precedence are sometimes necessary in cases of mixed
associativity.
For instance, the C++ expression \texttt{x.y()} requires that
``\texttt{.}'' bind tighter than ``\texttt{()}'', but this should not prevent \texttt{x().y}
from being a legal expression. A naive implementation, however, would stipulate that the
left-hand side of a dot should have precedence at least as high as the dot expression itself,
and thus would rule out the expression \texttt{x().y}.
With separate left-precedence and right-precedence, we avoid this issue:~a prefix operator only
constrains the right-precedence of its operands, while a postfix operator only constrains
the left-precedence.

\subsection{Recursive-descent actions}
\label{sec:rd}

One reason that hand-written recursive-descent parsers are often regarded as more intuitive
than generated LR parsers
is that it is usually clear, by examining the stack trace,
what the parser is ``doing'' at any point in time (viz., ``parsing an expression'', ``parsing an array literal'',
and so on).
By contrast, a typical LR DFA state may refer to multiple grammar rules,
and the intuitive meaning is often unclear
(e.g., $X \ra A \cdot B$ and $B \ra \cdot\,C D$ are connected by an $\eps$-edge in the NFA,
and hence end up grouped into the same vertex in the DFA).
In this section, we observe that we can recover much of the intuitive simplicity of recursive-descent
parsing within the LR paradigm, by introducing a new class of actions
for LR automata.
First, we assume that some of the occurrences
of nonterminals in the right-hand side of grammar rules
are designated as ``unfoldable'', and if an occurrence is {\em not}
unfoldable,
then it must be predicted at the beginning of its occurrence in the input, in recursive-descent style.
To make this distinction more precise, we describe the following modification to the LR~NFA.
(For simplicity, we describe only the case of $k = 0$, and omit the $k$-follow sets; the extension to
$k > 0$ is straightforward and follows that of Construction~\ref{cons:lrknfa}.)

\begin{construction}[RD NFA]\normalfont
\label{cons:rdknfa}
Fix a grammar $\GG$.
We define the RD($0$) NFA of $\GG$ as follows.
\begin{enumerate}
    \item The vertices consist of pairs $(\Pi,\,R)$,
        where $\Pi$ is 
        either a dotted production of $\GG$ or the special production
        $\#R \ra \cdot\,R$,
        and $R \in \Lambda \cup \{ \bot \}$
        is the nonterminal currently in left-recursive position
        (or $\bot$ if none).
    \item The starting vertex is $(\#S \ra \cdot\,S,\,S)$
        where $S$ is the start symbol of $\GG$.
    \item The edges are determined as follows:
    \begin{enumerate}
        \item Prediction edges: for every vertex $v = (X \ra \alpha \cdot Y \beta,\,R)$,
            if either of the following criteria hold:
            \begin{itemize}
                \item $Y$ is designated unfoldable in the production $X \ra \alpha \cdot Y \beta$, or:
                \item $Y = R$ and $\alpha = \eps$, i.e., the production is left-recursive
                    on $R$ (including the special production $\#R \ra \cdot\,R$);
            \end{itemize}
            then for every vertex $w = (Y \ra \cdot\,\gamma,\,R)$,
            then there is an $\eps$-edge from $v$ to $w$.
        \item Recur-step edges: for vertices $v = (X \ra \alpha \cdot Y \beta,\,R)$
            which do not meet the criteria of item 3(a),
            there is an edge with label $\mr{RecurStep}(Y)$
            from $v$ to $(\#Y \ra \cdot \, Y,\,Y)$.
        \item Step edges: for vertices $v = (X \ra \alpha \cdot \tau \beta,\,R)$
            and $w = (X \ra \alpha \,\tau \cdot \beta,\,\bot)$,
            there is an edge from $v$ to $w$ with label $\tau$
            (where $\tau$ is a single symbol, either terminal or nonterminal).
    \end{enumerate}
    \item The vertex accept actions are determined as follows:
    \begin{enumerate}
        \item For vertices $(X \ra \alpha \, \cdot\,,\,-)$,
            where $X$ is not the special symbol $\#R$,
            the associated action is ``Reduce $X \ra \alpha$''.
        \item For vertices $(\#R \ra R \, \cdot,\, -)$,
            the associated action is ``Return''.
        \item For vertices $(X \ra \alpha \cdot \sigma \beta,\,-)$,
            where $\sigma \in \Sigma$ is a terminal symbol,
            the associated action is ``Shift''.
        \item For vertices $v = (X \ra \alpha \cdot Y \beta,\,-)$
            which do not meet the criteria of item 3(a),
            the associated action is ``Recur~$Y$''.
    \end{enumerate}
\end{enumerate}
\end{construction}

The RD DFA is defined, as usual, by applying the standard NFA/DFA subset construction
to the RD NFA.
We note that while it is possible for the RD automaton to have additional conflicts
which the LR automaton would not
(i.e., when some items conflict with the new ``Recur''/``Return'' actions),
we can always avoid such conflicts in general by declaring the symbols in question
to be unfoldable in their respective productions.
In the extreme case, in which every symbol is unfoldable in every production,
we recover the standard LR behavior.

We now present the modified stack-based LR parsing algorithm
for such an RD DFA,
an adaptation of the standard LR algorithm~(Construction~\ref{cons:stacklrparse}).

\begin{construction}[Stack-Based RD Parsing Algorithm]\normalfont
    Fix a grammar $\GG$ and an integer $k$
    and let $D$ be a conflict-free RD($k$)~DFA
    for $\GG$.
    Let $S_0$ denote the starting state of~$D$,
    and fix an input string $x = x_1\ldots x_n$.
    The stack-based RD parsing algorithm on $x$ then operates as follows.
    \begin{enumerate}
        \item Initialize the state stack with a single element, $S_0$,
            and initialize the syntax stack to be empty.
            Initialize the cursor position $i = 1$.
        \item Repeat:
        \begin{enumerate}
            \item If the state stack is empty, halt with failure.
            \item Let $v = (\Pi_v, R)$ be the vertex on top of the state stack.
                Examine the current lookahead $\lambda$, i.e., the first $k$ symbols of
                $x_i x_{i+1} \ldots x_n \srt^k$.
                If $v$ has no action on $\lambda$, then halt with failure.
                Otherwise, act according to the action on $\lambda$:
                \begin{itemize}
                    \item If the action is ``Shift'',
                        then set the current buffer symbol to $x_i$,
                        push $x_i$ onto the syntax stack,
                        and increment $i$
                        (unless $i = n+1$, in which case halt with failure).
                    \item If the action is ``Reduce $\Pi = X \ra \alpha_1 \ldots \alpha_s$''
                        then pop $s$ elements off of the syntax stack,
                        construct a syntax element labeled ``$\Pi$'' with those $s$ arguments
                        as children,
                        and push the resulting element back onto the syntax stack.
                        Then, pop $s$ elements off of the state stack,
                        and set the current buffer symbol to $X$.
                    \item If the action is ``Recur $Y$'',
                        then set the current buffer symbol to $\mr{RecurStep}(Y)$.
                    \item If the action is ``Return'',
                        then pop $2$ elements off of the state stack
                        (i.e., $(\#R \ra R\,\cdot,\,R)$ and $(\#R \ra \cdot\,R,\,R)$),
                        and set the current buffer symbol to $R$.
                \end{itemize}
            \item Let $\alpha$ denote the current buffer symbol.
                \begin{itemize}
                    \item If $\alpha = S$, then if $i = n+1$, halt with success
                        and return the top (i.e., only) element of the syntax stack;
                        otherwise, halt with failure.
                    \item If $\alpha \neq S$,
                        then let $v'$ be the vertex for which there is an edge $v \overset{\alpha}{\ra} v'$
                        (there is exactly one such $v'$, since $D$ is a DFA).
                        Push $v'$ onto the state stack.
                \end{itemize}
        \end{enumerate}
    \end{enumerate}
\end{construction}

We also make one additional observation, concerning mutual left-recursion.
In our definition of the RD NFA~(Construction~\ref{cons:rdknfa}),
we have specified in the criteria of item 3(a)
that symbols which are immediately left-recursive
are to be unfolded automatically.
This is motivated by a number of common constructs,
e.g., left-associative operators
in operator-precedence expressions,
where it would be clearly invalid to make an explicit Recur call
(lest it trigger infinite recursion),
but where left-recursion is nonetheless desired.
However, there may be cases of mutual recursion (e.g., $A \ra B \ldots$ and $B \ra A \ldots\,$,
where neither $A$ nor $B$ is marked unfoldable.
This necessitates an additional check on the RD DFA:
if any vertex is self-reachable via RecurStep edges,
we must reject the grammar, as there exist input strings which may trigger infinite recursion.

\subsection{Parallel shift/reduce parsing (XLR)}
\label{sec:xlr}

Our final modification to the LR paradigm is a very general mechanism,
which we refer to as XLR (i.e., extended LR),
which, when all else fails, permits LR parsers to ``fork'' parallel copies
which can make different shift/reduce decisions in order to deal with conflicts.
We emphasize that XLR should only be used as a last resort---we
have found that when a parser requires XLR
(as opposed to LR with our extensions),
this generally indicates that parses are truly confusing for the human reader,
and the grammar should be modified to avoid this.
Nevertheless, we include a description
of the XLR paradigm here for completeness.

To begin with, we give a definition of the XLR family of grammars,
with LR as a special case.
\begin{definition}[LR Parse]\normalfont
Let $T \in \Sigma \cup \Lambda$ be a symbol of the grammar $\GG$,
and $\mc{P}$ be a parse tree for $T$ on some input string $x = x_1 \ldots x_n$.
An {\em LR parse} of $x$ for $T$, according to the parse tree $\mc{P}$,
consists of a sequence $\mc{S}$ of operations ``$\mr{Shift}(\sigma)$''
or ``$\mr{Reduce}(\Pi)$''
(where $\Pi$ is a production of~$\GG$),
which describes the items encountered in a postorder traversal of $\mc{P}$.
\end{definition}

\begin{definition}[Partial LR Parse]\normalfont
    \label{def:partparse}
Let $T \in \Sigma \cup \Lambda$ be a symbol of the grammar $\GG$,
and let 
$x = x_1 \ldots x_s$ be an input string.
A {\em partial LR parse} of $x$ for $T$ with lookahead $\lambda \in \Gamma^k$
consists of a sequence $\mc{S}_{\mr{part}}$
such that for some suffix $x_{s+1} \ldots x_n$,
with $\lambda$ matching the first $k$ symbols of $x_{s+1} \ldots x_n \srt^k$,
and some sequence $\mc{S}$ an LR parse of $x$ for $T$,
$\mc{S}_{\mr{part}}$ is the longest prefix of $\mc{S}$ that includes exactly $s$
``Shift'' operations.
\end{definition}

\noindent When it is clear from context, we omit the symbol $T \in \Sigma \cup \Lambda$,
and assume $T = S$, the starting symbol of the grammar $\GG$.

\begin{definition}[XLR]\normalfont
    \label{def:xlr}
For integers $k,t \geq 0$, we say that a grammar $\GG$ is XLR($k, t$)
if for every string $x$
and every lookahead $\lambda \in \Gamma^k$,
$x$ has at most $t$ distinct partial LR parses
with lookahead $\lambda$.
\end{definition}

Evidently, a grammar $\GG$ is XLR($k, 1$) if and only if it is LR($k$).
Moreover, we can adapt the standard stack-based LR parsing algorithm
to support LR($k$) automata with conflicts, via nondeterministic branching,
enabling us to parse the XLR($k, t$) grammars efficiently.

\begin{construction}[Stack-Based XLR Parsing Algorithm]\normalfont
    \label{cons:stackxlrparse}
    Fix a grammar $\GG$ and an integer $k$
    and let $D$ be the canonical LR($k$)~DFA for $\GG$
    (not necessarily conflict-free).
    Let $S_0$ denote the starting state of~$D$,
    and fix an input string $x = x_1\ldots x_n$.
    The stack-based XLR parsing algorithm on $x$ then operates as follows.
    \begin{enumerate}
        \item Initialize the state stack with a single element, $S_0$,
            and initialize the syntax stack to be empty.
            Initialize the cursor position $i = 1$.
        \item Repeat:
        \begin{enumerate}
            \item If the state stack is empty, halt with failure.
            \item Let $v$ be the vertex on top of the state stack.
                Examine the current lookahead $\lambda$, i.e., the first $k$ symbols of
                $x_i x_{i+1} \ldots x_n \srt^k$.
                If $v$ has no action on $\lambda$, then halt with failure.
                Otherwise, act according to the action or actions on $\lambda$,
                choosing among them nondeterministically\footnote{Of course,
                we can simulate this on a deterministic machine by
                forking multiple copies of the stacks,
                and executing all extant (non-halted) copies in parallel,
                symbol-by-symbol on the input string.} if there are multiple
                actions on $\lambda$:
                \begin{itemize}
                    \item If the action is ``Shift'',
                        then set the current buffer symbol to $x_i$,
                        push $x_i$ onto the syntax stack,
                        and increment $i$
                        (unless $i = n+1$, in which case halt with failure).
                    \item If the action is ``Reduce $\Pi = X \ra \alpha_1 \ldots \alpha_s$''
                        then pop $s$ elements off of the syntax stack,
                        construct a syntax element labeled ``$\Pi$'' with those $s$ arguments
                        as children,
                        and push the resulting element back onto the syntax stack.
                        Then, pop $s$ elements off of the state stack,
                        and set the current buffer symbol to $X$.
                \end{itemize}
            \item Let $\alpha$ denote the current buffer symbol.
                \begin{itemize}
                    \item If $\alpha = S$, then if $i = n+1$, halt with success
                        and return the top (i.e., only) element of the syntax stack;
                        otherwise, halt with failure.
                    \item If $\alpha \neq S$,
                        then let $v'$ be the vertex for which there is an edge $v \overset{\alpha}{\ra} v'$
                        (there is exactly one such $v'$, since $D$ is a DFA).
                        Push $v'$ onto the state stack.
                \end{itemize}
        \end{enumerate}
    \end{enumerate}
    At the end of execution, if all runs halt with failure,
    then the entire process halts with failure;
    if multiple runs halt with success,
    then the entire process halts with failure (i.e., ambiguity);
    and finally, if exactly one run halts with success,
    then the entire process halts with success and returns the parse
    resulting from the successful run.
\end{construction}

In order to establish the correctness of the XLR parsing algorithm,
we first describe an alternate perspective on the execution of LR automata:~specifically,
rather than performing an LR parse in the standard manner via the LR($k$)~DFA,
we observe that it can also be done (albeit perhaps not efficiently), nondeterministically, via the
NFA.
\begin{construction}[Nondeterministic LR Parsing Algorithm]\normalfont
    \label{cons:nondetlrparse}
    Fix a grammar $\GG$ and an integer $k$
    and let $N$ be the canonical LR($k$)~NFA (Construction~\ref{cons:lrknfa})
    for $\GG$.
    Let $S_0$ denote the starting state of~$N$,
    and fix an input string $x = x_1\ldots x_n$.
    The nondeterministic LR parsing algorithm on $x$ then operates as follows.
    \begin{enumerate}
        \item Initialize the state stack with a single element, $S_0$,
            and initialize the syntax stack to be empty.
            Initialize the cursor position $i = 1$.
        \item Repeat:
        \begin{enumerate}
            \item If the state stack is empty, halt with failure.
            \item Let $v$ be the vertex on top of the state stack.
                Nondeterministically execute one of the following actions:
                \begin{itemize}
                    \item Choose a vertex $w$ for which there exists an $\eps$-edge
                        $v \overset{\eps}{\ra} w$,
                        and push $w$ onto the state stack.
                    \item Examine the current lookahead $\lambda$, i.e., the first $k$ symbols of
                        $x_i x_{i+1} \ldots x_n \srt^k$.
                        If $v$ has no action on $\lambda$, then halt with failure.
                        Otherwise, nondeterministically choose one of the actions on $\lambda$:
                        \begin{itemize}
                            \item If the action is ``Shift'',
                                then set the current buffer symbol to $x_i$,
                                push $x_i$ onto the syntax stack,
                                and increment $i$
                                (unless $i = n+1$, in which case halt with failure).
                            \item If the action is ``Reduce $\Pi = X \ra \alpha_1 \ldots \alpha_s$''
                                then pop $s$ elements off of the syntax stack,
                                construct a syntax element labeled ``$\Pi$'' with those $s$ arguments
                                as children,
                                and push the resulting element back onto the syntax stack.
                                Then, pop $(s+1)$ elements off of the state stack,
                                and set the current buffer symbol to $X$.
                        \end{itemize}
                        Let $\alpha$ denote the current buffer symbol.
                        \begin{itemize}
                            \item If $\alpha = S$, then if $i = n+1$, halt with success
                                and return the top (i.e., only) element of the syntax stack;
                                otherwise, halt with failure.
                            \item If $\alpha \neq S$,
                                then let $v'$ be the vertex for which there is an edge $v \overset{\alpha}{\ra} v'$
                                (or, if none exists, halt with failure).
                                Push $v'$ onto the state stack.
                        \end{itemize}
                    \end{itemize}
        \end{enumerate}
    \end{enumerate}
    At the end of execution, if all runs halt with failure,
    then the entire process halts with failure;
    if multiple runs halt with success,
    then the entire process halts with failure (i.e., ambiguity);
    and finally, if exactly one run halts with success,
    then the entire process halts with success and returns the parse
    resulting from the successful run.
\end{construction}

We also introduce some notation for state stacks of the above algorithms.
Specifically, we write:
\[ T = T_0 \mapsto_{\tau_1} \ldots \mapsto_{\tau_m} T_m \]
to denote a stack $T$ of the XLR algorithm
with DFA vertices $T_0, \ldots, T_m$,
such that $\tau_i \in \Sigma \cup \Lambda$ is the label which caused $T_i$ to be pushed onto the stack.
Similarly, we write:
\[ U = U_0 \mapsto_{\sigma_1} \ldots \mapsto_{\sigma_r} U_r \]
to denote a stack $U$ of the nondeterministic LR algorithm
with NFA vertices $U_0, \ldots, U_r$,
such that $\sigma_i \in \Sigma \cup \Lambda \cup \{\eps\}$
is the label which caused $U_i$ to be pushed onto the stack.
We restrict our attention to {\em valid} stacks,
i.e., those for which each edge described in the stack actually exists in the corresponding automaton.

\begin{definition}[Related Stacks]\normalfont
    Let $T$ be a state stack of the XLR algorithm,
    and let $U$ be a state stack of the nondeterministic LR algorithm.
    We say $T$ is {\em related} to $U$,
    denoted $T \triangleleft U$,
    if the sequence of symbols for $T$ is identical to the sequence
    of non-$\eps$ symbols for $U$.
\end{definition}

\begin{lemma}\normalfont
    \label{lem:parsecompl}
    Let $\sigma_1 \cdots \sigma_r \in (\Sigma \cup \Lambda)^*$ be a sequence of symbols,
    let $x_1 \cdots x_s \in \Sigma^*$ be an input string,
    let $\lambda \in \Gamma^k$ be a lookahead, and let
    $U = U_0 \mapsto_{\sigma_1} \ldots \mapsto_{\sigma_r} U_r$
    be a valid stack of the nondeterministic LR algorithm such that $U_r$ has an action on $\lambda$.
    If $\sigma_1 \cdots \sigma_r \genstar x_1 \cdots x_s$ with shift/reduce sequence $\mc{S}_{\mr{part}}$,
    then there exists $x_{s+1} \cdots x_n \in \Sigma^*$ with $\fst_k(x_{s+1} \cdots x_n \srt^k) = \lambda$
    such that $S \genstar x_1 \cdots x_n$ with shift/reduce sequence $\mc{S}$,
    where $\mc{S}_{\mr{part}}$ is a prefix of $\mc{S}$.
\end{lemma}
\begin{proof}
    As in the conflict tracing algorithm,
    letting $U_i = (X_i \to \alpha_i \cdot \beta_i, \{\mu_i\})$,
    we construct the tail string $\beta_r \ldots \beta_0$,
    then choose $x_{s+1} \cdots x_n$ such that $\beta_r \ldots \beta_0 \genstar x_{s+1} \cdots x_n$
    and $\fst_k(x_{s+1} \cdots x_n \srt^k) = \lambda$
    (the existence of such $x_{s+1} \cdots x_n$
    follows from the correctness of the LR($k$) NFA construction).
    By a standard induction on the derivation, such a string takes the nondeterministic LR algorithm to completion.
\end{proof}

\begin{lemma}\normalfont
    \label{lem:parttfae}
    Let $x = x_1 \ldots x_n$ be an input string,
    let $s \in \{0,\ldots,n\}$,
    and let $\lambda$ be the lookahead for the partial input $x_1\ldots x_s$
    (i.e., the first $k$ symbols of $x_{s+1}\ldots x_n \srt^k$).
    Then the following are equivalent:
    \begin{enumerate}
        \item $\mc{S}_{\mr{part}}$ is a partial LR parse of $x_1\ldots x_s$
            with lookahead $\lambda$ (Definition~\ref{def:partparse}).
        \item For some run of the nondeterministic LR parsing algorithm
            (Construction~\ref{cons:nondetlrparse}) on $x$, with cursor at $s$,
            such that there is an action from the current state on lookahead $\lambda$,
            the sequence of shift/reduce operations that have been performed
            is $\mc{S}_{\mr{part}}$.
        \item For some run of the XLR parsing algorithm (Construction~\ref{cons:stackxlrparse}) on $x$,
            with cursor at $s$, such that there is an action from the current state on lookahead $\lambda$,
            the sequence of shift/reduce operations that have been performed
            is $\mc{S}_{\mr{part}}$.
    \end{enumerate}
\end{lemma}
\begin{proof}
    To show that (2) implies (1),
    we note that by correctness of the shift/reduce operations performed
    in the nondeterministic LR parsing algorithm,
    we have $\sigma_1 \cdots \sigma_r \genstar x_1 \ldots x_s$
    with reduction sequence $\mc{S}_{\mr{part}}$,
    where $\sigma_1 \cdots \sigma_r$ are the labels on the stack.
    The claim then follows from Lemma~\ref{lem:parsecompl}.

    To show that (3) implies (1),
    we note that by correctness of the shift/reduce operations performed
    in the XLR parsing algorithm,
    we have $\sigma_1 \cdots \sigma_r \genstar x_1 \ldots x_s$.
    Moreover, by correctness of the NFA/DFA subset construction,
    for a given valid stack $T$ of the XLR algorithm,
    there exists a valid stack $U$ of the
    nondeterministic LR algorithm, with $T \triangleleft U$,
    and with the same action on lookahead $\lambda$.
    The claim then follows again from Lemma~\ref{lem:parsecompl}.

    To show that (1) implies (2),
    we observe that there is a run of the nondeterministic LR parsing algorithm
    that corresponds to any given partial parse,
    namely,
    one which nondeterministically decides to predict each production via an $\eps$-edge
    precisely when its occurrence begins in the input.

    Finally, to show that (2) implies (3), fix a run of the nondeterministic LR algorithm,
    and construct a run of the XLR algorithm as follows.
    We maintain the invariants, inductively, that the state stack $T$ of the XLR algorithm
    is related to the state stack $U$ of the nondeterministic LR algorithm
    ($T \triangleleft U$),
    and that the cursor of the XLR algorithm is at the same point as that of the
    nondeterministic LR algorithm.
    \begin{itemize}
        \item When the nondeterministic LR algorithm makes an $\eps$-edge prediction,
            do nothing.
            This preserves relatedness of the stacks, since we do not count $\eps$-edges
            in the path in the stack~$U$.
        \item When the nondeterministic LR algorithm performs a ``Shift'' operation,
            perform a ``Shift'' operation.
            Such an action is permitted at the current state by the correctness of the
            NFA/DFA subset construction (since the stacks are related),
            and the cursor advances in both cases.
        \item When the nondeterministic LR algorithm performs a ``Reduce'' operation
            by a production~$\Pi$,
            perform a ``Reduce'' operation by the same production $\Pi$.
            Such an action is permitted at the current state by the correctness of the
            NFA/DFA subset construction (since the stacks are related).
            Moreover, by construction, the top of the NFA stack must consist of vertices
            $(X \ra \alpha \cdot Y \beta, -) \mapsto_{\eps} (Y \ra \cdot\,\sigma_1 \ldots \sigma_s, -) \mapsto_{\sigma_1} \ldots \mapsto_{\sigma_n} (Y \ra \sigma_1 \ldots \sigma_s\,\cdot, -)$.
            Since the stacks are related (and the XLR stack cannot contain $\eps$-edges),
            popping $(s+1)$ items from the nondeterministic LR stack,
            and popping $s$ items from the XLR stack,
            preserves relatedness of the stacks, as desired. \qedhere
    \end{itemize}
\end{proof}

\begin{theorem}[Efficient XLR Parsing]\normalfont
    For fixed $k, t \geq 0$, the XLR($k, t$) grammars can be parsed in linear time.
\end{theorem}
\begin{proof}
    It suffices to show that Construction~\ref{cons:stackxlrparse}
    runs in linear time.
    This, in turn, will follow if we show that at most $t$ parallel copies
    of the execution are extant (non-halted) at any point in the input string.
    Suppose we have some $t'$ parallel copies, with cursor $s$,
    each with an action on the current lookahead $\lambda$
    (this is without loss of generality, since
    a copy with no action on $\lambda$ halts immediately with failure).
    Each such copy has a distinct sequence of shift/reduce operations.
    Moreover, by Lemma~\ref{lem:parttfae},
    the operation sequence of each such copy
    determines a distinct partial parse of the input string
    with the given lookahead $\lambda$.
    Since $\GG$ is XLR($k, t$), we conclude that $t' \leq t$,
    as desired.
\end{proof}

We remark that the XLR parsing algorithm is similar to generalized LR (GLR) parsing algorithms
such as the ETL family, in that it allows for multiple partial parses
to coexist simultaneously.
It differs, however, in that the partial parses do not share state
(e.g., reusing parses of substrings).
This means that for some (non-XLR) grammars, XLR parsing may require exponential time,
while the dynamic-programming approaches of ETL will always require at most cubic time.
On the other hand, it also means that the constant factor on the running time of XLR
parsing is very small, and does not depend on $|\GG|$.
As a result, for small $t$,
XLR($k, t$) parsing remains competitive with hand-written recursive-descent parsers,
while algorithms such as ETL incur significant overhead.

Of course, the XLR algorithm would be of limited practical utility
if we could not determine whether grammars of interest are XLR($k, t$).
Naively, Definition~\ref{def:xlr} seems quite intractable,
as it quantifies over all input strings $x$.  
However, there are many possible heuristics that are effective in practice.
We give one such approach here.
(For simplicity, we confine our attention to the canonical LR($k$) automaton;
the extension to the optimized automata of Section~\ref{sec:optaut} is straightforward.)

\begin{definition}[Fork Point]\normalfont
    Let $N$ be the canonical LR($k$)~NFA for the grammar $\GG$.
    Suppose there is an LR conflict on some vertices $w_1,\ldots,w_d$,
    i.e., these vertices are each reachable
    from the starting vertex on some common string $\tau \in (\Lambda \cup \Sigma)^*$,
    and have pairwise distinct actions on some common lookahead $\lambda \in \Gamma^k$.
    We say that a vertex $v$ is a {\em fork point} (of degree $d$) for the conflict
    if every path from the starting vertex to one of $w_1,\ldots,w_d$
    passes through $v$ exactly once,
    and the outgoing $\eps$-edges $v \overset{\eps}{\ra} x$ can be partitioned into $d$ sets $S_1,\ldots,S_d$
    such that all paths
    from the starting vertex to $w_i$ pass through an edge in $S_i$.
\end{definition}

\begin{definition}[Join-Safe Vertex]\normalfont
    \label{def:joinsafevert}
    Let $N$ be the canonical LR($k$)~NFA for the grammar $\GG$,
    and let $v$ be a vertex of $N$.
    For an outgoing $\eps$-edge $v \overset{\eps}{\ra} w$,
    if $w = (X \ra \cdot\,\eta, -)$,
    define the {\em join tail} of the edge to be the string $\eta$.
    We say that $v$ is {\em join-safe}
    if for every pair of paths $v \overset{\eps}{\ra} w_1 \ra \ldots \ra x_1$,
    $v \overset{\eps}{\ra} w_2 \ra \ldots \ra x_2$
    on the same string $\tau$ such that $x_1,x_2$ have distinct actions on some
    common lookahead $\lambda \in \Gamma^k$,
    if $\delta_1$ is the join tail of $v \overset{\eps}{\ra} w_1$
    and $\delta_2$ is the join tail of $v \overset{\eps}{\ra} w_2$,
    then there do not exist $z,z' \in \Sigma^*$ such that $\delta_1 \genstar z$
    and $\delta_2 \genstar zz'$;
    and similarly there do not exist $z,z' \in \Sigma^*$
    such that $\delta_2 \genstar z$
    and $\delta_1 \genstar zz'$.
\end{definition}

We remark that although the criterion of Definition~\ref{def:joinsafevert}
is still undecidable in general,
in practical cases it tends to be a relatively straightforward theorem-proving task.

\begin{definition}[Fork Degree]\normalfont
    Let $N$ be the canonical LR($k$)~NFA for the grammar $\GG$,
    and suppose that every LR conflict has a fork point that is join-safe.
    We define the {\em fork degree} of $N$ to be the
    minimum value, over assignments of LR conflicts each to one of their respective join-safe fork points,
    of the maximum value,
    over all paths $v_1 \ra \ldots \ra v_n$ where $v_1$ is the starting vertex,
    of the product of the maximum fork degrees of each of $v_1,\ldots,v_n$,
    where we define the fork degree of a non-fork point to be $1$.
    (Note that if a fork point always occurs on a reachable cycle in $N$,
    irrespective of which assignment is chosen,
    then the fork degree of $N$ is $\infty$.)
\end{definition}

\begin{theorem}[XLR Fork Bound]\normalfont
    \label{thm:xlrforkbound}
    Let $N$ be the canonical LR($k$)~NFA for the grammar $\GG$,
    and suppose $N$ has finite fork degree $t$.
    Then $\GG$ is XLR($k, t$).
\end{theorem}
\begin{proof}
    Fix an assignment of LR conflicts to one of their respective join-safe fork points.
    Partition the partial runs of the nondeterministic LR parsing algorithm
    into equivalence classes by the sequence of reduce operations that they have performed
    (indexed by $i$ as in Definition~\ref{def:partparse}).
    Now, when a given partition forks
    (i.e., multiple runs within the partition have distinct shift/reduce actions on some lookahead),
    since the runs have made the same shift/reduce decisions thus far,
    there must be a single string $\tau \in (\Lambda \cup \Sigma)^*$
    that takes the NFA starting state to a vertex with each of the distinct actions.
    By assumption, there is some corresponding fork point $v = (X \ra \alpha \cdot Y \beta, -)$
    such that for some vertices
    $w_1 = (Y \ra \cdot\,\eta_1, -)$, $\ldots$\,, $w_d = (Y \ra \cdot\, \eta_d, -)$,
    the respective stacks contain the vertices $(v, w_j)$, sequentially, for each $j$.

    Since no $w_j$ can be popped from its respective stack until a string generated
    by $\eta_j$
    has subsequently been seen in the input (and, until such a point, the substring being parsed
    must be a prefix of one generated by $\eta_j$),
    and since $v$ is join-safe (Definition~\ref{def:joinsafevert}),
    we conclude that if any partition $j$ later forks on a vertex
    not reachable from $v$ in the NFA,
    then no other partition $j' \neq j$ can be extant at that time.
    Hence, all nontrivial forks occur at successive points on paths in the NFA.
    Since distinct partial parses require distinct shift/reduce decisions in the
    nondeterministic LR parsing algorithm (Lemma~\ref{lem:parttfae}),
    the claim then follows by definition of XLR($k, t$).
\end{proof}

Many other heuristics to prove the XLR criterion are possible;
for instance, we may consider refining the join-safety criterion
to specify that a vertex may be deemed safe if no conflicts are reachable
once its parent vertices have been advanced by its predicted symbol.
However, the above suffices to prove grammars XLR in a wide variety of practical cases.
As an example, we consider the following simple (non-LR($1$)) grammar:
\[ S' \ra S \]
\vspace*{-1.75em}
\[ S \ra X\,c\ a \ |\ Y\,c\ b \]
\[ X \ra c \]
\[ Y \ra c \]
There is a single conflict in the LR($1$)~DFA,
namely, at the vertex $\{ (X \ra c \,\cdot, \{c\}), (Y \ra c \,\cdot, \{c\}) \}$
on lookahead~$c$.
However, this conflict has the fork point $(S' \ra \cdot\,S, \{\srt\})$,
which has $\eps$-edges to the two vertices $(S \ra \cdot\,X\,c\ a, \{\srt\})$
and $(S \ra \cdot\,Y\,c\ b, \{\srt\})$.
Since clearly no instance of $X\,c\ a$ can occur as a prefix of an instance of $Y\,c\ b$,
and vice versa,
the fork point is join-safe,
and we conclude that the grammar is XLR($1, 2$), as expected.

\section{Benchmarks}

To test the generality of the new paradigms proposed in this work,
we investigate the possibility of using them to parse full
industrial programming languages.
Some care needs to be taken in selecting the languages to prototype.
For instance, it is known that parsing C++ is undecidable---the classic
example being the simple pointer/operator ambiguity:
\[ \text{\texttt{x * y;}} \]
Whether this is parsed as a declaration (``\texttt{y} is a pointer
of type \texttt{x*}'') or a multiplication expression
(``\texttt{x} times \texttt{y}'')
depends on whether \texttt{x} names a type or a value,
but this may depend on arbitrary template computation,
which is known to be Turing-complete.
More generally, a language construct need not be literally undecidable
in order to be a poor example;
it suffices that it be truly confusing for a human observer.

With this in mind, we have selected two well-specified industrial
languages: Python 3.9.12 and Golang 1.17.8.
We have expressed each language's syntax in a form amenable
to our LR software implementation,
and have endeavored the match the language specifications
as closely as possible, with the following caveats:
\begin{itemize}
    \item Both languages' specifications include some provision
        to limit the set of valid Unicode character encodings
        (e.g., to restrict
        characters used in identifiers to code points
        designated as ``letters'').
        Enumerating these encodings is quite laborious
        and depends on the details of the language's compiler implementation,
        and so we have opted not to enforce these restrictions,
        instead allowing arbitrary Unicode
        code points (except those that are otherwise designated
        as distinguished by the language).
        This does not interfere with the parsing task,
        and should it be desired to enforce a particular
        set of restrictions,
        this can easily be done as a postprocessing pass
        on the abstract syntax tree.
    \item In the case of Python specifically,
        there are several properties which are designated
        as ``syntax errors'' by the compiler,
        but which often involve some semantic complexity
        (e.g., exception clauses must have either
        at least one ``\texttt{except}'' or at least one ``\texttt{finally}'',
        and the default ``\texttt{except}'' must appear in last position).
        While we believe it would be possible to implement
        such restrictions using our attribute mechanism
        (Section~\ref{sec:attr}),
        this would greatly complicate the grammar.
        Moreover, we believe that LR attributes should be used only
        when absolutely necessary to resolve ambiguities
        or conflicts---if parsing is possible without
        encoding semantic properties as attributes,
        this should be preferred, as the desired semantic properties
        can be easily enforced via postprocessing
        on the abstract syntax tree.
        Thus, we have opted not to enforce this type of
        restriction at the grammar level,
        making our Python parser accept a language that is
        slightly more general than the official specification.
\end{itemize}
\noindent
Subject to these caveats,
we have successfully generated parsers
for both of the above industrial languages,
and tested it on the standard language implementation codebases
(i.e., the directories ``\texttt{cpython/Lib}'' and ``\texttt{go/src}'',
respectively, from the official language repositories, exempting ``bad'' test
examples which the standard compilers also fail to parse).
Tests are performed on an AMD Ryzen Threadripper PRO 3995WX CPU
with 256~GiB of main memory, running Ubuntu Linux 22.04.
The results are as follows:
\begin{itemize}
    \item For Golang 1.17.8, our grammar specification requires
        551 lines of code,
        and parser generation runs in 61 sec,
        producing 69 KLOC of generated code.
        On the standard codebase described above
        (parsing each file 10 times),
        the standard parser requires 18.28 sec,
        and our generated parser requires 15.24 sec
        (1.2x faster).
    \item For Python 3.9.12, our grammar specification requires
        398 lines of code,
        and parser generation runs in 48 sec,
        producing 70 KLOC of generated code.
        On the standard codebase described above
        (parsing each file 10 times),
        the standard parser requires 33.64 sec,
        and our generated parser requires 7.78 sec
        (4.3x faster).
\end{itemize}
We remark that our generated parsing code is heavily optimized
(and includes, among various other enhancements, a pool-based
memory allocator); however, the parser generation code itself
is not significantly optimized,
and there is plenty of room should its
running time need to be improved.
The above benchmarks serve as a proof-of-concept
that, using the techniques described
in this work, automatic LR parser generation is a practical option
for industrial programming language grammars.

\ \\
\noindent \textbf{Addendum.} In the interim since this work
was undertaken, both Golang and Python have introduced
modifications to their respective languages
that make them much more difficult to parse:
\begin{itemize}
    \item Golang 1.18 introduced generic types,
        supporting declarations such as the following:
        \[ \texttt{type x [T interface\{\}] struct \{ y T \} } \]
        Generics are a valuable language feature at the semantic level.
        Unfortunately, the syntactic implementation creates
        new opportunities for confusing parses
        such as the following:
        \[ \texttt{type x [a *b] c} \]
        similar to the classic C++ ambiguity,
        where it is unclear if we are parsing a generic declaration
        with type parameter \texttt{a} of type \texttt{*b},
        or a non-generic declaration of an array of type
        \texttt{[a*b]c}, i.e., a length of \texttt{a*b} elements each of type \texttt{c}.
        The Golang specification indicates that this ambiguity
        is to be resolved by preferring to parse as an array type
        whenever the tokens enclosed by the brackets could
        be parsed as a valid expression.
        Unfortunately, such a rule is not compositional:~in order to write
        a BNF-style specification, we would need to
        include a criterion such as ``a type parameter
        consists of an input sequence which is {\em not} parseable
        as an expression''.
        Moreover, even this criterion is not known until the end of
        the bracketed expression,
        which may require the parser to make decisions with an unbounded
        amount of lookahead.
    \item Python 3.10, via PEP 634, introduced structural pattern-matching,
        supporting statements such as the following:
        \[ \hspace{-7em} \texttt{match x:} \] \vspace{-1.8em}
        \[ \hspace{-2em} \texttt{case x0,x1:} \] \vspace{-1.8em}
        \[ \hspace{-1em} \texttt{y = x0} \] \vspace{-1.8em}
        \[ \hspace{-0.4em} \texttt{case x0,x1,x2:} \] \vspace{-1.8em}
        \[ \hspace{-0.9em} \texttt{y = x2} \]
    Though seemingly innocuous from a parsing perspective,
    the specification of PEP 634 also requires that \texttt{match}
    become a ``soft keyword'', i.e., that it can still be used
    as an identifier in addition to being used as a keyword.
    This creates the potential for confusing partial parses such as the following:
        \[ \texttt{match(x, y, *z, *w} \ldots \]
    where it is unclear if we are matching on a tuple
    which has inline list expansions \texttt{z},\texttt{w},
    or alternatively if we are calling the identifier
    ``\texttt{match}'' with arguments \texttt{x}, \texttt{y}, \texttt{*z}
    and \texttt{*w}.
    If, for instance, we had written \texttt{**w} in place of \texttt{*w},
    the compiler would be required to flag this as an error
    if we are parsing a pattern-match,
    but not if we are parsing a function call---but we
    will not know which is the case until we reach
    the end of the line, which could be an unbounded number of tokens away.
\end{itemize}

\noindent While we believe that some of these language modifications could be handled
in practice via the XLR parsing algorithm~(Section~\ref{sec:xlr}),
we believe it is also the case that these new features
result in partial parses which are truly confusing for a human reader,
and hence fall outside the scope of the present work.

\section{Conclusion}

Parsing is a laborious and error-prone component of compiler construction,
and despite known theoretical approaches to automatic parser generation,
most industrial parsers are still written by hand.
In this work, we have demonstrated that automatic parser generation can in fact be practical,
by introducing a number of new developments upon the standard LR parsing paradigm of Knuth~et~al.
With our new methodology,
we can automatically generate efficient parsers for virtually all programming languages
that are intuitively ``easy to parse''---a claim we have supported experimentally,
by implementing the new algorithms
and running them on syntax specifications for Golang and Python~3.
We believe that with the introduction of this and related work,
parsing will no longer be a significant barrier to programming language development.

\section{Acknowledgements}

The author would like to thank George Kulakowski and Gideon Wald for many helpful
discussions and comments on earlier drafts of this work.

\bibliographystyle{alpha}
\bibliography{parsing}

\end{document}